\documentclass[aps,amssymb,twocolumn,superscriptaddress,amsmath,latexsym, tightenlines,nofootinbib]{revtex4}

\usepackage{setspace}
\usepackage{graphicx}
\usepackage{psfrag}

\newcommand{\bie}{\begin{itemize}}
\newcommand{\eie}{\end{itemize}}

\newcommand{\scr}{\scriptstyle}
\newcommand{\noi}{\noindent}
\newcommand{\nn}{\nonumber}

\newcommand{\lab}{\label}

\newcommand{\beq}{\begin{equation}}
\newcommand{\eeq}{\end{equation}}
\newcommand{\bea}{\begin{eqnarray}}
\newcommand{\eea}{\end{eqnarray}}
\newcommand{\ba}{\begin{array}}
\newcommand{\ea}{\end{array}}

\newcommand{\na}{\nabla}
\newcommand{\pa}{\partial}
\newcommand{\eq}{\equiv}
\newcommand{\fr}{\frac}
\newcommand{\sr}{\sqrt}
\newcommand{\ha}{\fr{1}{2}}

\newcommand{\al}{\alpha}
\newcommand{\bt}{\beta}
\newcommand{\ga}{\gamma}
\newcommand{\de}{\delta}
\newcommand{\ep}{\epsilon}

\newcommand{\te}{\theta}

\newcommand{\la}{\lambda}

\newcommand{\si}{\sigma}

\newcommand{\ph}{\phi}
\newcommand{\vp}{\varphi}
\newcommand{\ch}{\chi}
\newcommand{\ps}{\psi}
\newcommand{\om}{\omega}

\newcommand{\De}{\Delta}

\newcommand{\Ph}{\Phi}
\newcommand{\Ps}{\Psi}
\newcommand{\Om}{\Omega}

\newcommand{\krr}{K^r{}_r}
\newcommand{\ktt}{K^\te{}_\te}
\newcommand{\rtoi}{r\to\infty}

\newcommand{\nj}{^{^{\scr n}}_{_{\scr j}}}
\newcommand{\no}{^{^{\scr n}}_{_{\scr 1}}}
\newcommand{\npoj}{^{^{\scr n+1}}_{_{\scr j}}}

\newcommand{\njpo}{^{^{\scr n}}_{_{\scr j+1}}}
\newcommand{\njmo}{^{^{\scr n}}_{_{\scr j-1}}}
\newcommand{\njpt}{^{^{\scr n}}_{_{\scr j+2}}}
\newcommand{\njmt}{^{^{\scr n}}_{_{\scr j-2}}}
\newcommand{\njph}{^{^{\scr n}}_{_{\scr j+\ha}}}
\newcommand{\njmh}{^{^{\scr n}}_{_{\scr j-\ha}}}
 
\newcommand{\Dtp}{\De^{^{\scr t}}_{_{\scr +}}}
\newcommand{\Ddiss}{\De^{^{\scr t}}_{_{\scr +{\rm KO}}}}

\newcommand{\Drz}{\De^{^{\scr r}}_{_{\scr 0}}} 
\newcommand{\Drp}{\De^{^{\scr r}}_{_{\scr +}}}
\newcommand{\Drm}{\De^{^{\scr r}}_{_{\scr -}}}
\newcommand{\Drzb}{\De^{^{\scr r}}_{_{\scr 0 b}}}
\newcommand{\Dhrz}{\De^{^{\scr \fr{r}{2}}}_{_{\scr 0}}}
\newcommand{\Drsz}{\De^{^{\scr r^2}}_{_{\scr 0}}} 
\newcommand{\Drcz}{\De^{^{\scr r^3}}_{_{\scr 0}}} 
\newcommand{\Dhrcz}{\De^{^{\scr \fr{r^3}{2}}}_{_{\scr 0}}} 

\newcommand{\mutp}{\mu^{^{\scr t}}_{_{\scr +}}}

\newcommand{\murm}{\mu^{^{\scr r}}_{_{\scr -}}}
\newcommand{\bmurp}{\bar{\mu}^{^{\scr r}}_{_{\scr +}}}
\newcommand{\bmurm}{\bar{\mu}^{^{\scr r}}_{_{\scr -}}}
\begin{document}

\bibliographystyle{unsrt}

\title{Final Fate of Subcritical Evolutions of Boson Stars}

\preprint{AEI-2007-135}

\author{Chi Wai Lai}
	\email{cwlai@ust.hk}
	\affiliation{
		Dept.~of Physics and Astronomy, University of British Columbia, Vancouver BC, V6T 1Z1 Canada
	}
	\affiliation{
		Dept.~of Physics, The Hong Kong University of Science and Technology, Clear Water Bay, 
		Kowloon, Hong Kong
	}
\author{Matthew W.\ Choptuik}
	\email{choptuik@physics.ubc.ca}
	\affiliation{
		Dept.~of Physics and Astronomy, University of British Columbia, Vancouver BC, V6T 1Z1 Canada
	}
	\affiliation{
		CIFAR Cosmology and Gravity Program
	}
	\affiliation{
		Max-Planck-Institut f\"ur Gravitationsphysic, Albert-Einstein-Institut, Am M\"uhlenberg 1, D-14476 Golm, Germany
	}


\begin{abstract}
We present results from a study of Type I critical phenomena in the dynamics 
of general relativistic boson stars in spherical symmetry. The boson stars 
are modelled with a minimally coupled, massive complex field (with no 
explicit self-interaction), and are driven to the threshold of black hole 
formation via their gravitational interaction with an initially imploding 
pulse of massless scalar field.  Using a distinct coordinate system, we reproduce 
previous results~\cite{shawley:phd,scott_matt:2000}, including the scaling 
of the lifetime of near-critical configurations, as well as the fact that such
configurations are well described as perturbed, one-mode-unstable boson stars.
In addition, we make a detailed study of the long-time evolution of marginally 
subcritical configurations. Contrary to previous 
claims~\cite{shawley:phd,scott_matt:2000}, we find that the end state in
such cases does not involve dispersal of the bulk of the boson star field to 
large radial distances, but instead can be generically described by a 
stable boson star executing large amplitude oscillations.  Furthermore
we show that these oscillations can be largely identified as excitations of 
the fundamental mode associated with the final boson star, as computed in
perturbation theory.
\end{abstract}

\maketitle

\section{Introduction}
Over the past decade or so, intricate and unexpected phenomena related to 
black holes have been discovered through the detailed numerical 
study of various models for gravitational collapse,
starting with one of the authors' investigation of the spherically symmetric 
collapse of a massless scalar field~\cite{choptuik}.
These studies generally concern the {\em threshold} of black hole 
formation (a concept described below), and the phenomena observed
near threshold are collectively called (black hole) critical phenomena,
since they share many of the features associated with critical phenomena
in statistical mechanical systems.  The study of critical phenomena 
continues to be an active area of research in numerical relativity, 
and we refer the interested reader to the review article
by Gundlach~\cite{gundlach:2003} for full details on the subject. 
Here we will simply summarize some key points that are most germane 
to the work described in this paper.

To understand black hole critical phenomena, one must understand
the notion of the ``threshold of black hole formation".  
The basic idea is to consider {\em families} of solutions of
the coupled dynamical equations for the gravitational field 
and the matter field that is undergoing collapse (a complex 
scalar field, $\phi$, in our case).  Since we are considering a dynamical problem,
and since we assume that the overall dynamics is uniquely determined 
by the initial conditions, we can view the families as being 
parametrized by the initial conditions---variations in one or more 
of the parameters that fix the initial values will then generate 
various solution families.  We also emphasize that we are considering 
collapse problems.  This means that we will generically 
be studying the dynamics of systems that have length scales 
comparable to their Schwarzschild radii, {\em for at least some period of time 
during the dynamical evolution}.  We also note that we will often
take advantage of the complete freedom we have as numerical
experimentalists to choose initial conditions that lead to collapse, but which may 
be highly unlikely to occur in an astrophysical setting.

We now focus attention on single parameter families of 
data, so that the specification of the initial data is fixed 
up to the value of {\em the} family parameter, $p$.  We will generally 
view $p$ as a non-linear control parameter that will
be used to govern how strong the gravitational field becomes
in the subsequent evolution of the initial data, and in particular,
whether a black hole forms or not. Specifically, we will always 
demand that any one-parameter family of solutions has the 
following properties:
\begin{enumerate}  
\item For sufficiently small values of $p$ the dynamics remain regular 
      for all time, and no black hole forms.  
\item For sufficiently large values of $p$, complete gravitational collapse 
      sets in at some point during the dynamical development of the initial
      data, and a black hole forms.
\end{enumerate}
From the point of view of simulation, it turns out to be a relatively 
easy task for many models of collapse to construct such families,
and then to identify two specific parameter values, $p^-$ ($p^+$) which do not (do)
lead to black hole formation.  Once such a  ``bracket'' $[p^-,p^+]$ has been
found, it is straightforward in principle to use a technique such as 
binary search to hone in on a {\em critical parameter value}, $p^\star$, such 
that all solutions with $p<p^\star$ ($p>p^\star$) do not (do) contain
black holes. A solution corresponding to $p=p^\star$ thus sits at the 
threshold of black hole formation, and is known as a {\em critical solution}.
It should be emphasized that underlying the existence of critical solutions 
are the facts that (1) the end states (infinite-time behaviour)  corresponding to
properties 1~and 2~above are {\em distinct} (a spacetime containing a black hole 
{\em vs} a spacetime not containing a black hole) and (2) the process 
characterizing the black hole threshold (i.e.\ gravitational collapse) 
is {\em unstable}. We also note that we will term evolutions with $p<p^\star$
{\em subcritical}, while those with $p>p^\star$ will be called {\em supercritical}.

Having discussed the basic concepts underlying black hole critical phenomena,
we now briefly describe the features of critical collapse that are most 
relevant to the research described below.

First, critical solutions {\em do}
exist for all matter models that have been studied to date, and for 
any given matter model, almost certainly constitute discrete sets.  In
fact, for some models, there may be only {\em one} critical solution,
and we therefore have a form of universality.  

Second, critical solutions 
tend to have additional symmetry beyond that which has been adopted in the 
specification of the model (e.g. we will impose spherical 
symmetry in our calculations).  

Third, the critical solutions known
thus far, and the black hole thresholds associated with them, come 
in two broad classes.  The first, dubbed Type I, is characterized 
by static or periodic critical solutions (i.e.\ the additional symmetry 
is a continuous or discrete time-translational symmetry), and by
the fact that the black hole mass just above threshold is {\em finite}
(i.e.\ so that there is a minimum black hole mass that can be 
formed from the collapse). 
The second class, called Type II, is characterized by continuously or 
discretely self-similar critical solutions (i.e.\ the additional
symmetry is a continuous or discrete scaling symmetry), and by the 
fact that the black hole mass just above threshold is {\em infinitesimal}
(i.e.\ so that there is {\em no} minimum for the black hole mass that 
can be formed).    The nomenclature Type I and Type II is by analogy 
with first and second order phase transitions in statistical mechanics,
with the black hole mass viewed as an order parameter.

Fourth, solutions close to criticality exhibit various scaling laws.
For example, in the case of Type I collapse, where the critical solution is an
unstable, time-independent (or periodic) compact object, the amount 
of time, $\tau$, that the dynamically evolved configuration is well 
approximated by the critical solution {\em per se} satisfies a scaling law of 
the form
\beq
	\label{tau-scaling}
	\tau(p) \sim -\gamma \ln | p - p^\star | \,,
\eeq
where $\gamma$ is a {\em universal} exponent in the sense of not 
depending on which particular family of initial data is used to 
generate the critical solution, and $\sim$ indicates that the relation 
(\ref{tau-scaling}) is expected to hold in the limit $p \to p^\star$.

Fifth, and finally, much insight into critical phenomena comes 
from the observation that although unstable, critical solutions 
tend to be {\em minimally} unstable, in the sense that they 
tend to have only a few, and perhaps only one, unstable modes 
in perturbation theory.  In fact, if one assumes that a Type I
solution, for example, has only a single unstable mode, then 
the growth factor (Lyapunov exponent) associated with that mode can 
be immediately related to the scaling exponent~$\gamma$ defined by
(\ref{tau-scaling}).

In this paper we will be exclusively concerned with Type I critical 
phenomena, where the threshold solutions will generally turn
out to be unstable boson stars.  Previous work relevant to ours
includes studies by Hawley \cite{shawley:phd} and Hawley \& Choptuik~\cite{scott_matt:2000} of 
boson stars in spherically symmetry.
We extend this work and show that, contrary to 
previous claims~\cite{shawley:phd,scott_matt:2000}
 that subcritical solutions disperse most of 
the original mass of the boson star to large distances---the late time behaviour of subcritical 
evolution is characterized by oscillation about a stable boson star solution.
We also apply a linear perturbation analysis similar to
that in~\cite{shawley:phd,scott_matt:2000} and confirm that the observed oscillation modes agree with
the fundamental modes given by perturbation theory.  (We use a code 
kindly provided by S.~Hawley \cite{hawley:private} to
generate the frequencies from the perturbation analysis.)  


The outline of the rest of this paper is as follows: in Section~\ref{sec:mathform}
we describe the mathematical formulation for our numerical simulations, which
includes ~\ref{sec:model}: the model for the boson stars, and \ref{sec:IVP}: the 
initial value problem.  In Section~\ref{sec:results} we present results of our
simulations: in~\ref{sec:setup} we present the setup of numerical experiments, 
in~\ref{sec:criticalphenomena} the Type I character of the critical
solutions is demonstrated, \ref{sec:latetime} contains a discussion of the end state of
subcritical evolutions and is followed by some perturbation analysis in~\ref{perttheory}.
Section~\ref{sec:conclusions} summarizes our findings, while the finite difference
approximations used and our convergence testing of our implementations of them, 
are given in Appendices~\ref{sec:FDA} and \ref{sec:convergencetest} respectively.  

In what follows 
we base our work in the context of classical field theory, and 
we choose units in which $G=c=1$.  In addition, without loss of generality, we restrict
ourselves to the case where the particle mass, $m$, associated with the complex 
scalar field satisfies $m=1$.

\section{Mathematical Formulation}
\lab{sec:mathform}
\subsection{The model}
\lab{sec:model}
Our model for boson stars involves a self-gravitating massive 
complex scalar field, $\ph = \ph_1 + i \ph_2$, minimally coupled to 
gravity as given by general relativity.  (Note that we do not make the complex scalar 
field explicitly self-interacting---in the literature, the stationary
 configurations 
in this case are sometimes called ``mini'' boson stars.) An additional, massless real 
scalar field, $\ph_3$, also minimally coupled to gravity, is used to dynamically
``perturb" the boson star.  The interaction between the massive complex 
scalar field and the massless real scalar field is thus through the gravitational 
field alone.
The whole system can be described by the action 

\bea
 S = \int d^4 x \sr{-g} \left[ \fr{R}{16 \pi}\right.  &-&  \left. \ha \left(\na^\mu \ph
\na_{\mu} \ph^* + m^2 \ph \ph^*  \right) \right. \nonumber\\
   &-& \left. \ha \na^{\mu} \ph_3 \na_{\mu}
\ph_3 \right] \,,
\eea

\noi
where $R$ is the spacetime Ricci scalar and $m$ is the  mass 
of the bosonic particle.
Variations of the action with respect to the metric, $g_{\mu \nu}$, the complex
scalar field, $\ph$, and the real scalar field, $\ph_3$,
yield
the equations of motion, which are the Einstein equation, the Klein-Gordon 
equation and the wave equation, respectively: 

\beq \lab{eintein_eq}
 R_{\mu \nu} - \ha g_{\mu \nu} R = 8 \pi T_{\mu \nu}\,,
\eeq

\beq \lab{klein_gordon_eq}
\na^{\mu}\na_\mu \ph - m^2 \ph   = 0\,,
\eeq

\noi
and 

\beq
\na^{\mu}\na_\mu \phi_3 = 0\,,
\eeq

\noi
where

\beq
T_{\mu \nu} = T_{\mu \nu}^\phi + T_{\mu \nu}^{\phi_3}\,,
\eeq

\bea \lab{Ctmunu}
 T^\phi_{\mu \nu} &\eq&
\ha \left[  \left( \na_{\mu} \ph \na_{\nu} \ph^\ast + \na_{\nu} \ph
\na_{\mu} \ph^\ast \right)
\right. \nonumber \\ & & \quad\quad
\left. - g_{\mu \nu} \left( \ \na^\al \ph \na_\al \ph^\ast + m^2 |\ph|^2
\right)\right] \,,
\eea

\beq \lab{tmunu_r}
T_{\mu \nu}^{\phi_3} = 
  \na_{\mu} \phi_3 \na_{\nu} \phi_3
- \ha g_{\mu \nu}  \ \na^\al \phi_3 \na_\al \phi_3 \,.
\eeq

Equations (\ref{eintein_eq})--(\ref{tmunu_r}) completely determine the
dynamics of our system (up to coordinate transformations), 
once appropriate
initial conditions and boundary conditions are specified.

To study the system numerically, we adopt the standard ``3+1" ADM
formalism~\cite{ADM:1962,york:1978}. Since we restrict ourselves to 
spherically symmetry
the metric can be written in a much simpler form than in the generic case. 
Here we
use maximal-isotropic coordinates, which is a different system than 
that used 
in~\cite{shawley:phd,scott_matt:2000}.  
We note in passing that 
although the accuracy of finite difference calculations in any
given coordinate system can in principle be estimated using intrinsic means (e.g.\
convergence tests), we feel that it is nonetheless useful to reproduce the calculations of
\cite{shawley:phd,scott_matt:2000} in a distinct coordinate system.  

In maximal-isotropic coordinates, the line element can be written as: 

\bea \lab{metric}
  d s^2 = \left( -\al^2 + \ps^4 \bt^2 \right) dt^2 &+& 2 \ps^4 \bt\, dt\,
dr \nonumber \\ &+& \ps^4 \left( dr^2 + r^2 d \Om^2 \right) \,,
\eea

\noi
where $\al$, $\bt$ and $\ps$ are the lapse function, $r$-component of the
shift vector and the conformal factor respectively, and all are  functions
of $t$ and $r$.  We further define new variables to transform the Klein
Gordon and wave equations into a first order (in time) system:

\beq
  \Ph_i \eq \ph_i'\,, 
\eeq

\beq
  \Pi_i \eq \fr{\ps^2}{\al} \left( \dot{\ph_i } - \bt \ph_i'\right)\,,
\eeq

\noi
where $i = 1, 2$ or $3$, $'\eq \pa/\pa r$ and $\dot{} \eq \pa / \pa t$.
As with the geometric variables, $\phi_i$. $\Phi_i$ and $\Pi_i$ are 
functions of $t$ and $r$ alone.

With these definitions, the 
Hamiltonian constraint and momentum constraints are given
by~\cite{cwlai:phd}

\bea \lab{hamiltonian_constraint}
\fr{3}{\ps^5} \fr{d}{dr^3} &&\!\!\!\!\!\!\!\!\!\!\left( r^2 \fr{d \ps}{dr} \right) + \fr{3}{16}
{\krr}^2 
= \nonumber \\
&-& \pi\left( \fr{ \sum_{i=1}^3 \left( \Ph^2_i + \Pi^2_i\right)  }{\ps^4} +
m^2 \sum_{i=1}^2 {\ph_i}^2 \right) \!,
\eea

\beq \lab{momentum_constraint}
 {\krr}\left.'\right. + 3 \fr{( r \ps^2)'}{r \ps^2} {\krr} = -\fr{8
\pi}{\ps^2} \left( \sum_{i=1}^3 \Pi_i  \Ph_i \right) \,,
\eeq

\noi
and the Klein-Gordon and wave equations become

\bea \lab{klein_gordon_eq1}
 \dot{\ph}_i & = & \fr{ \al}{\ps^2} \Pi_i + \bt \Ph_i \,, \\
 \dot{\Ph}_i & = & \left( \bt \Ph_i + \fr{\al}{\ps^2} \Pi_i\right)' \,, \\ \nn
 \dot{\Pi}_i & = & \fr{3}{\ps^4}  \fr{d}{d r^3}\left[ r^2 \ps^4 \left( \bt
\Pi_i + \fr{ \al }{ \ps^2} \Ph_i\right)\right]- \al \ps^2 m^2 \ph_i
\left( 1 - \de_{i3} \right) \\ \lab{klein_gordon_eq3}
&& -  \left( \al K^{r}\,_{r} + 2 \bt \fr{(r \ps^2)'}{r \ps^2} \right) \Pi_i
\,.
 \eea

In addition to equations (\ref{hamiltonian_constraint})--(\ref{klein_gordon_eq3}), 
we need to determine the lapse function and shift component using our specific coordinate
choices.
The maximal
condition, which maximizes the 3-volume of the $t={\rm const.}$ slices, is given by
$K \equiv K^i\,_i = 0$.  This is implemented 
by choosing initial data so that $K(0,r)\equiv0$ and 
then demanding that 
\beq
 \dot{K}(t,r) = 0\,.
\eeq
\noi
for all $t$ and $r$.

This leads to the following linear ODE for $\alpha(t,r)$, which must be solved at each instant
in time in order to maintain the maximal condition~\cite{cwlai:phd}:

\bea \lab{maximal_condition}
  \al'' &+& \fr{2}{r \ps^2} \fr{ d}{dr^2}\left( r^2 \ps^2 \right) \al' 
  \nonumber\\ &+& \left( 4 \pi
m^2 \ps^4 \sum_{i=1}^2 \ph_i^2 - 8 \pi \sum_{i=1}^3 \Pi_i^2 - \fr{3}{2} (\ps^2
K^r\,_r)^2 \right) \al  \nonumber \\ &=&  0\,.
\eea

The
isotropic condition, which is implicit in the chosen form of the 
metric~(\ref{metric}), demands that the 3-metric of each $t={\rm const.}$ hypersurface be 
conformally flat.  This requirement leads to the following ODE for the shift vector 
component, $\beta(t,r)$:

\beq \lab{isotropic_condition}
 r \left( \fr{\bt}{r}\right)' = \fr{3}{2} \al K^r\,_r\,.
\eeq

Equations (\ref{hamiltonian_constraint})--(\ref{isotropic_condition})
constitute a complete set of differential equations governing our model.  
Note that our approach is an instance of so-called fully 
constrained evolution, wherein all of the  geometric
variables---$\ps$ and $K^r\,_r$ in this case---are computed at each time step using 
constraint equations.
To completely fix a solution---given initial data---we also must impose regularity 
and boundary conditions at $r=0$ and $r\to\infty$ respectively.

Regularity at the origin, $r=0$, requires
\bea
 \ps'(t,0) &=& 0\,, \\
 \krr(t,0) &=& 0 \,, \\
 \al'(t,0) &=& 0\,, \\
 \ph_i'(t,0) &=& 0\,, \\
 \Pi_i'(t,0) &=& 0\,,
\eea

\noi
whereas the outer boundary conditions are

\beq \lab{psbc1}
\lim_{r\to\infty}\ps(t,r) = 1 + \fr{C(t)}{r} + O(r^{-2}) \,,
\eeq

\bea
\lim_{r\to\infty}\al(t,r) &=& \lim_{r\to\infty}\fr{2}{\ps(t,r)}-1 \nonumber \\ 
  &=& 1 - \fr{2C(t)}{r} + O(r^{-2}) \,,
\eea

\beq
\lim_{r\to\infty}\bt(t,r) = \fr{D(t)}{r} + O(r^{-2}) \,,
\eeq

\noi
and

\bea
 \dot{\Ph_i}+ \Ph_i'+\fr{\Ph_i}{r} &=& 0\,, \\ \lab{last_BC}
 \dot{\Pi_i}+ \Pi_i'+\fr{\Pi_i}{r} &=& 0\,,
\eea

\noi
for some functions $C(t)$ and $D(t)$.  The last two of these equations are approximate 
Sommerfeld conditions that assume that as $r\to\infty$, the three scalar field components, 
$\phi_i$, are purely outgoing with amplitudes decaying as $1/r$.
For given initial data, eqs.~(\ref{hamiltonian_constraint})--(\ref{last_BC}) 
now completely determine our system.

For diagnostic purposes, we also define the mass aspect function

\beq 
 M(t,r) \eq \left( \fr{ \ps^2 r}{2} \right)^3 K^r\,_r\,^2 - 2 \ps' r^2 \left( \ps
+ r \ps'\right) \,,
\eeq

\noi
which is equal to the ADM mass in any vacuum region exterior to the support
of matter.  


In addition, although $\ps$ and $\krr$ are ultimately determined from
the constraint equations,
the following evolution equations are used for providing initial estimates
for the iterative constraint-solving process.
%
\beq  \lab{ps_evolution_eq}
 \dot{\ps} = -\ha \al \ps \krr + \fr{\left( \ps^2 \bt\right)'}{2 \ps} \;,
\eeq

\bea \lab{krr_evolution_eq}
  \dot{\krr} = \bt \krr' &-& \fr{2 \al}{(r \ps^2)^2} + \fr{2}{r^2 \ps^6}
\left[ \al r \left( r \ps^2\right)'\right]' \nonumber \\
  &+& 8 \pi m^2 \al |\ph|^2\;.
\eea

Full details of our finite differencing scheme are given in App.~\ref{fda}.

\subsection{The initial value problem}
\lab{sec:IVP}
The primitive object in our model problem is a (ground-state) boson star, represented
by a configuration of massive complex scalar field, centred at the origin.  
Ideally one would like a ``star'' to be  
described by a localized, time-independent matter source that generates an
everywhere regular (i.e.\ non-singular) gravitational field.
However, for the case of a complex scalar field, it can be shown that such
regular, time-independent configurations do not exist~\cite{Friedberg:1987}.  Despite 
this fact, since the stress-energy tensor 
(\ref{Ctmunu}) depends only on the modulus of the scalar field (and the 
gradients of the modulus), one {\em can} construct scalar field configurations with harmonic 
time-dependence that produce time-independent metrics.  
Specifically, we adopt the following ansatz for boson
stars in spherical symmetry:

\beq \lab{ansatz}
 \ph(t,r) = \ph_0(r)\, e^{-i \om t}\,,
\eeq

\noi
and then demand that the spacetime be static, i.e.\ we demand that the metric admits a timelike
Killing vector field, $\ch$, which is orthogonal to the $t={\rm const.}$ surfaces.
Adapting coordinate time to the 
timelike Killing vector field, we have
\beq
 \bt = 0\,, 
\eeq
\noi
for all time $t$.  Additionally, we have that the time derivatives of any of the geometrical
variables identically vanish.  It then follows immediately
that~\cite{cwlai:phd}
\beq
 \krr = 0\,.
\eeq

As is necessary for the consistency of the ansatz~(\ref{ansatz}), the 
isotropic condition for $\beta$~(\ref{isotropic_condition}) 
is automatically satisfied, and we are left with geometrical variables 
$\al(0,r)$, $\ps(0,r)$ and $\ph_0(r)$
that need to be determined from the maximal slicing condition (\ref{maximal_condition}),
the Hamiltonian constraint (\ref{hamiltonian_constraint}) and the Klein-Gordon equation (\ref{klein_gordon_eq3}),
respectively:

\bea \lab{bs1divp1}
 \ps' & = & \Ps \,,\\
 \Ps' & = & -  \fr{ 2\Ps}{r} - \pi \left[ \ps \Ph^2 + \ps^5 \left( \fr{
\om^2}{\al^2} + m^2 \right) \ph^2 \right] \,,\\
  \ph' & = & \Ph \,,\\
  \Ph' & = & - \left( \fr{ 2}{r} + \fr{ A}{\al} +  \fr{ 2\Ps}{\ps} \right)
\Ph + \ps^4  \left( m^2 - \fr{ \om^2}{\al^2}\right) \ph \,,\\
  \al' & = & A \,,\\\lab{bs1divp6}
 A' & = & - 2 \left( \fr{ 1}{r} + \fr{ \Ps}{\ps}\right) A + 4 \pi \ps^4
\al \left( \fr{ 2 \om^2}{ \al^2} - m^2 \right) \ph^2\,.
\eea

Here, in order to simplify notation, we have dropped the subscript ``0'', making the 
identifications $\ph(r) \equiv \ph_0(r)$ and $\Ph(r) \equiv \ph'(r) \equiv \ph_0'(r)$.
We have also introduced auxiliary variables $\Psi(r) \equiv \psi'(r)$
and $A(r) \equiv \al'(r)$ in order to cast the above system of nonlinear ODEs in a 
canonical first-order form.
We assert that for any given value of $\ph(0) \equiv \ph_0(0)$,
the system~(\ref{bs1divp1})-(\ref{bs1divp6}) 
constitutes an eigenvalue problem with eigenvalue $\om = \om(\ph(0))$.  That is, for any specific 
value of $\ph(0)$ (which one can loosely view as being related to the central density of 
the star), a solution of~(\ref{ansatz}) that satisfies the appropriate regularity and 
boundary conditions will only exist for some specific 
value of $\om$.
The system~(\ref{bs1divp1})--(\ref{bs1divp6}) must be supplemented by boundary conditions, some of which 
are naturally applied at $r=0$, with the rest naturally set at $r=\infty$.  In particular,
regularity at $r=0$ implies 
\bea
\Ps(0) &=& 0\,, \\
\Ph(0) &=& 0 \,, \\
A(0) &=& 0 \,,
\eea
\noi
while at the outer boundary, we have
\bea \lab{psouterBC}
 \lim_{\rtoi}\ps(r) &=& 1 - \fr{C}{r} \,, \\ 
 \lim_{\rtoi}\ph(r) &\approx& 0\,,\\ \lab{alouterBC}
  \lim_{\rtoi}\al(r) &=& \fr{2}{\ps} - 1\,.
\label{al-obc}
\eea
\noi
Here the second condition follows from the expectation that $\ph$ should decay exponentially~\cite{kaup:1968} as $r\to0$.

We further note that due to the homogeneity and linearity of the slicing equation, we 
can always arbitrarily (and conveniently) choose the central value of the lapse via
\beq
 \al(0) = 1\,,
\eeq
\noi
and then, after integration of~(\ref{bs1divp1})-(\ref{bs1divp6}), can rescale $\al$ and $\om$ 
simultaneously to satisfy the true outer boundary condition, (\ref{al-obc}), for $\al$:
\bea
 \al(r) &\longrightarrow& c\, \al(r) \,,\\
 \om(r) &\longrightarrow& c\, \om(r) \,.
\eea
\noi
where $c$ is given by 
\beq
 c = \fr{2/\ps(r_{\rm max}) - 1}{\al(r_{\rm max})}\,,
\eeq
\noi
and $r_{\rm max}$ is the radial coordinate of the outer boundary of the 
computational domain.  

As mentioned above, any solution of~(\ref{bs1divp1})-(\ref{bs1divp6}) can
be conveniently labelled by the central value of the modulus of the scalar 
field, $\ph_0(0)=\ph(0)$.  For any given value of $\ph_0(0)$, we must then determine the 
eigenvalue, $\omega$, and in the current case of maximal-isotropic coordinates, the 
central value of the conformal factor $\psi(0)$, so that all of the boundary conditions 
are satisfied.  In principle, we can compute pairs $[\omega,\psi(0)]$ as a function
of $\ph_0(0)$ using a two-parameter ``shooting'' technique~\cite{numrec,jason:phd}.

Alternatively, in some cases 
we generate boson star initial data 
in maximal-isotropic coordinates by first constructing the stars in so-called 
polar-areal coordinates, and then performing a coordinate transformation on the 
resulting solution.

Polar-areal coordinates, which have seen widespread use in spherically symmetric 
computations in numerical relativity, can be viewed as the generalization of the 
usual Schwarzschild coordinates to {\em time-dependent}, spherically symmetric 
spacetimes. As with maximal slicing, the slicing condition in this case---known
as polar slicing---is expressed as a condition on the mean extrinsic curvature:
\beq
 K =\krr\,.
\eeq
\noi
Since in general we have $K = K^i{}_i = K^r{}_r + 2K^\theta{}_\theta$, this condition 
is implemented by requiring 
\beq
	K^\theta{}_\theta(t,r) = {\dot K^\theta{}_\theta}(t,r) = 0  \,,
\eeq
for all $t$ and $r$.

The spatial coordinates are fixed by demanding that the coordinate $r$ measure 
proper surface area (i.e.~that it be an {\em areal} coordinate).
It can be shown that this choice of $r$, together with polar slicing, further imply 
that $\beta\equiv0$, so that the line element becomes
\beq
 ds^2 = -\al^2 dt^2 + a^2 dr^2 + r^2 d \Om^2\,.
\eeq
\noi

As before, to construct star-like solutions, we adopt the time-harmonic ansatz~(\ref{ansatz})
for the complex scalar field,
adapt the time coordinate to the timelike Killing vector field, and require the
spacetime to be static.  We again find that the extrinsic curvature tensor vanishes 
identically (so that, for static data, the slicing is polar as well as maximal),
and that the momentum constraint~(\ref{momentum_constraint}) is automatically satisfied.

Again, considering the Hamiltonian constraint, the Klein-Gordon equation, and the slicing condition
\beq
\dot{\ktt} = 0\,,
\eeq
\noi
at $t=0$, we have (dropping the subscript 0's  as before):
\bea \lab{mcsfivpeq1}
  a' & = & \ha \Bigg\{ \fr{a}{r} \left( 1-a^2\right)\Bigg. \nonumber \\
  && \quad \,\, \left.  + \, 4 \pi r a \left[ \ph^2 a^2
\left( m^2 + \fr{\om^2}{\al^2}\right) + \Ph^2 \right] \right\} \,, \\
\al' & = & \fr{\al}{2}\Bigg\{ \fr{a^2-1}{r} \Bigg. \nonumber \\ 
  && \quad \,\, \left. + \, 4 \pi r \left[  a^2 \ph^2
 \left(  \fr{\om^2}{\al^2} - m^2 \right) 
  + \, \Ph^2 \right]\right\} \,, \\
  \ph' & = & \Ph \,, \\ \lab{mcsfivpeq4}
  \Ph' & = & - \left( 1+a^2 - 4 \pi  r^2 a^2 m^2\ph^2 \right) \fr{\Ph}{r} \nonumber\\
 &&- \left(  \fr{\om^2}{\al^2} - m^2\right) \ph a^2\,.
\eea
\noi
In this case, the regularity conditions are
\bea
 a(0) &=& 1 \, , \\
 \Ph(0) &=& 0\,,
\eea
\noi
while the outer boundary conditions are
\bea
 \lim_{\rtoi}\ph(r) &\approx & 0\,, \lab{phouterBC} \\
 \lim_{\rtoi}\alpha(r) &=& \frac{1}{a(r)} \lab{alphaobc}\,.
\eea
As before, we can convert the last condition to an {\em inner} condition on $\alpha$ by 
taking advantage of the linearity and homogeneity of the slicing equation.  Specifically, 
we can again choose $\alpha(0)=1$, and then after integration of 
(\ref{mcsfivpeq1})--(\ref{mcsfivpeq4})
simultaneously rescale $\al(r)$ as well as the eigenvalue, $\om$, so that 
(\ref{alphaobc}) is satisfied.  

We again consider the family of boson star solutions parametrized by the central
value of the modulus of the scalar field, $\ph_0(0)$.  In this case, given a value of 
$\ph_0(0)$, and using the conditions $a(0)=1$, $\alpha(0)=1$, $\Phi(0)=0$, we need only 
adjust the eigenvalue $\omega$ itself in order to generate a solution with the appropriate 
asymptotic behaviour (i.e.\ so that $\lim_{r\to\infty} \phi(r) = 0$). This is a classic
1-parameter shooting problem, which is comparatively easier than the 2-parameter shooting 
method described above.

Once we have computed a solution in areal coordinates, we can perform a coordinate
transformation from areal coordinates to isotropic coordinates \cite{dinverno,LPPT} (recall
that the maximal and polar slices coincide for the static case).
Essentially this amounts to solving an ODE of the form
\bea  \nn
 \left. r\,\right|_{R=R_\mathrm{max}} &=& \left[\left(
\fr{1+\sr{a}}{2}\right)^2 \fr{R}{a}\right]_{R=R_\mathrm{max}}\,, \\
 \lab{ODEb} \fr{dr}{dR} &=& \, a \fr{r}{R}\,.
\eea

We emphasize that 
(\ref{bs1divp1})--(\ref{bs1divp6}) or
(\ref{mcsfivpeq1})--(\ref{mcsfivpeq4}) are used for generating 
initial data describing a spacetime that has no matter content other than a single 
boson star.
To ``perturb'' a given boson star, and, in particular, to drive the star to the threshold
of black hole formation,
we implode a (spherical) shell of massless scalar
field onto it.  Specifically, we choose initial data for the massless field of the 
following ``gaussian'' form:
\beq
  \label{ph3-gaussian}
  \ph_3(0,r) = A_3 \, \exp \left[ - \left( \fr{ r - r_0}{\si} \right)^2 \right] \, ,
\eeq
where $A_3, r_0$ and $\si$ are adjustable parameters, controlling the overall 
amplitude, position and width, respectively, of the imploding gaussian wave packet.  To ensure that 
the massless field is almost purely in-going at the initial time, we specify the 
``conjugate'' variable $\Pi_3 \equiv \ps^2 / \alpha \left( \dot{\phi_3}  - \bt {\ph_3}'\right)$ as follows:
\beq
 \Pi_3(0,r) = -\left(\Ph_3(0,r) + \fr{\ph_3(0,r)}{r}\right) \, .
\eeq
In all of our studies described below, we have fixed $r_0$ and $\si$ in~(\ref{ph3-gaussian})
to $r_0 = 40$ and $\si=5$.  This ensures that the support of the massless field is well 
separated from that of the complex field (i.e.~from any of the boson stars {\em per se}
that we study) at the initial 
time.  

Once the complex scalar field, $\ph$, and the real scalar field,
$\ph_3$, are known, the initial data for the functions
$\ps(0,r)$, $\krr(0,r)$, $\al(0,r)$ and $\bt(0,r)$ are computed by solving
the Hamiltonian constraint (\ref{hamiltonian_constraint}), 
the momentum constraint (\ref{momentum_constraint}),
the slicing condition (\ref{maximal_condition}),
and the isotropic condition (\ref{isotropic_condition}) respectively.

\section{Results}
\lab{sec:results}

\subsection{Setup of numerical experiments}
\lab{sec:setup}

The PDEs solved in the simulations discussed here are those 
listed in the previous section.
We also provide a summary of the equations of motion of the system,
the boundary conditions, and details
of the finite difference approximation used in App.~\ref{sec:FDA}.
Additionally, results of convergence tests of our code are discussed in 
App.~\ref{sec:convergencetest}.

In order to study critical behaviour in the model we start with initial
data for the complex field that represents a boson star on the stable
branch (i.e.\ a star with a central scalar field value 
$\ph_0(0) \lesssim 0.08$,
using our units and conventions).
We generally choose a configuration that is reasonably relativistic,
i.e.\ with $\ph_0(0)$ bounded away from 0, but not too close to the
instability point, $\ph(0) \approx 0.08$.

Details of the initial data setup were described in the previous section.
Typical evolution of such initial data proceeds 
as follows.  Once we have fixed the boson star configuration, we complete
the specification of the massless scalar field initial data by fixing
the overall amplitude factor, $A_3$, and then evolve the system.
Initially, the shell of massless scalar field implodes towards $r=0$
at the speed of light, while the boson star ``sits'' in its static
state centered at the origin.  As the in-going massless shell reaches
the region of space occupied by the boson star, its contribution to the
overall gravitational field tends to compress the boson star to a higher
mean density and smaller radius.  The massless field passes through the
origin and then ``explodes'' outward, eventually propagating off the
computational domain.  Depending on the strength of the perturbation
from the massless field, we find that the compressed boson star either
relaxes to something resembling a stable boson star with large-amplitude
oscillations, or collapses to form a black hole.  Thus by adjusting
the massless scalar amplitude factor, $A_3$---which we generically use
as the adjustable parameter, $p$, in our study of critical behaviour
in the model---we can tune the evolution to the threshold of black
hole formation.   In practice we use a bisection search to refine our
estimate of the critical value, $A_3^\star$, and can carry the search
to machine precision, so that $\Delta A_3 / A_3
\sim 10^{-15}$ using standard 8-byte floating-point arithmetic.  
Unless otherwise specified, the computations described below have been 
performed with a spatial mesh spacing $\De r = 50/1024
\approx 0.049$, a Courant factor $\De t/\De r = 0.3$, and the
coefficient of Kreiss-Oliger dissipation, $\ep_d = 0.5$ (see App.~\ref{sec:FDA}
for the definition of $\ep_d$).  

We note that our numerical calculations generate entire {\em families} of 
critical solutions, fundamentally reflecting the fact that there is a
continuum of one-mode unstable boson star configurations (see Fig.~\ref{transit}).
In addition, for any fixed initial boson star state, the specifics of the 
observed threshold solution will depend on the details of the ``perturbing'' 
scalar field.  This last fact is, however, irrelevant to the conclusions 
that we draw from our study.

In the following section we discuss results from detailed studies of black
hole threshold solutions generated from several distinct initial boson
star states.  Table~I summarizes the values of $\phi_0(0)$
that were used, the approximate values of $A_3$ required to generate
a critical solution, the location, $r_{\rm max}$, of the outer boundary of the 
computational domain,
and the figures that display results associated
with the respective calculations.  Since we will not dwell on this
point below, we note that all of our calculations confirm the basic
picture previously reported that the black holes that form just above
threshold in this type of collapse generically have {\em finite} mass
(i.e.\ that the critical transition is Type I).

\vspace{0.5cm}
\begin{table*}[htbp]
\label{table-I}
\begin{center}
\begin{tabular}{| c | c | c | c |}
\hline
Fig.             & $\ph_0$ & $A^\star_3$ & $r_{\rm max}$\\ \hline
\hline
\ref{dmdr}       & 0.05       & 0.0032     &50 \\
\ref{modph}, \ref{transit}& 0.035, 0.04, 0.05     & 0.00471, 0.00342, 0.00316 &50\\
\ref{t_vs_lndp} & 0.02, 0.035, 0.04, 0.05 &  0.00915, 0.00471, 0.00342, 0.00316 &50\\
\ref{tmrphps} & 0.04 & 0.00342 &200\\ 
\ref{modphrmax100} &0.035, 0.04, 0.05 & 0.0083, 0.0061, 0.0031 &100\\
\ref{modph.04}, \ref{fft.04} &0.04 & 0.00342, 0.00603, 0.00623, 0.00632 &50, 100, 200, 400\\ 
\hline
\end{tabular}
\caption[Summary of parameters used to generate results displayed in Figs.~\ref{dmdr}-\ref{fft.04}]{
Summary of parameters used to generate the results displayed in Figs.~\ref{dmdr}-\ref{fft.04}.
Listed for each distinct computation or numerical experiment are 
the relevant figure numbers, central amplitude of the complex field, $\ph_0(0)$, 
the overall
massless scalar amplitude factor, $A_3^\star$ (see~\ref{ph3-gaussian}), 
that generates a marginally-critical solution, and the maximum radial coordinate, 
$r_{\rm max}$, of the computational domain. 
Other parameters defining the massless scalar initial profile~(\ref{ph3-gaussian})
are held fixed at $r_0 = 40$, $\si = 5$ for all simulations.  Other numerical parameters are
chosen to be $\De r = 50/1024 \approx 0.049$, $\De t/\De r = 0.3$ and $\ep_d = 0.5$, and 
are also fixed for the calculations discussed here.
}
\end{center}
\end{table*}

\begin{figure}
\begin{center}
\includegraphics[width=8.0cm,clip=true]{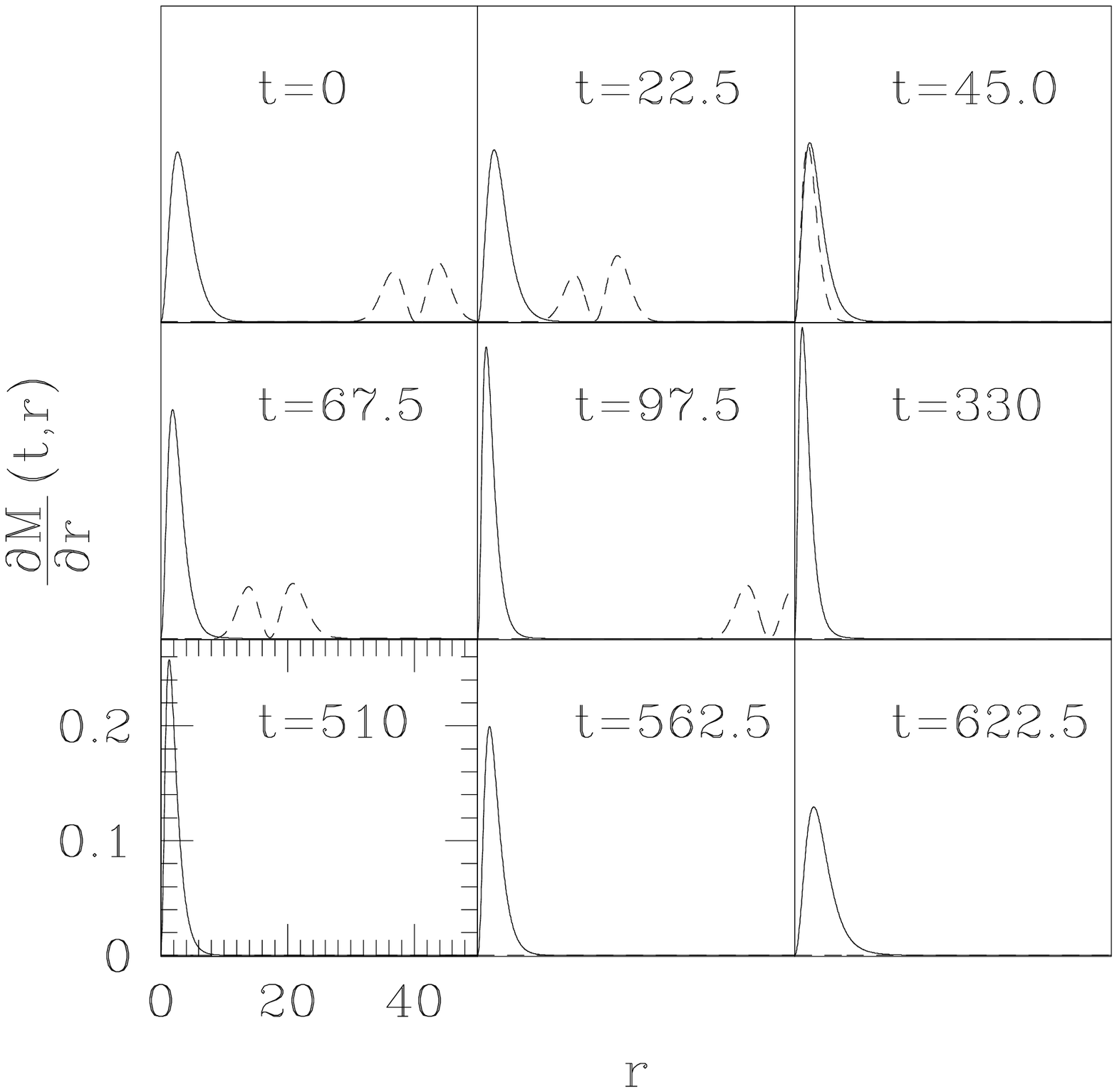}
\caption
[Critical evolution of a perturbed boson star with $\ph_0(0) = 0.05$
and mass $M_C = 0.62$]
{\small Critical evolution of a perturbed boson star with $\ph_0(0) = 0.05$
and mass, $M_{\rm ADM} = 0.62$ (using our units and conventions).  This figure shows the time development of 
contributions to $\pa M/\pa r$ from the complex (solid line) and real (dashed
line) scalar fields. 
Note that the temporal spacing between successive snapshots is {\em not} constant---the time instants
displayed have been chosen to illustrate the key features of the near-critical evolution.
Also note that we have multiplied the value of $\pa M/\pa r$
for the {\em real} scalar field by a factor of 8 to aid in the visualization of that 
field's dynamics.
The evolution begins with a stable boson star centered
at the origin, and an in-going gaussian pulse (shell) of massless, real scalar field 
that is used to perturb the star.  The overall amplitude factor, $A_3$,  of the initial real scalar 
field profile (see~(\ref{ph3-gaussian})), is the control 
parameter for generating the one-parameter family of solutions that interpolates through the 
black hole threshold. For the calculation shown
here, $A_3$ has been tuned to a critical value $A_3^\star \approx 0.0032$ via a bisection search (and 
with a fractional precision of $\approx 10^{-15}$).  
The other parameters defining the gaussian initial
profile of the massless field are $r_0 =40$ and $\si=5$.
The snapshots show that the real scalar field enters the region 
containing the bulk of boson star at $t\approx 22$, implodes through the origin
at $t \approx 45$, leaves the boson star region at $t\approx70$, and, finally, completely 
disperses from the computational domain at $t \approx 100$.  The boson star enters the critical
state at roughly the same time that the real field leaves the domain, and remains in that 
state for a period of time which is long compared to the crossing time of the massless field.
At $t\approx 510$, the boson star begins to depart significantly from the critical state.
}
\label{dmdr}
\end{center}
\end{figure}
\subsection{Critical phenomena}
\lab{sec:criticalphenomena}
We start by examining results from a critically perturbed boson star having an unperturbed central 
field value $\phi_0(0) = 0.05$.  As just described, the critical massless amplitude factor, 
$A_3^\star \sim 0.0032$, was determined by performing a bisection search on $A_3$, to roughly 
machine precision. (Recall that each iteration in this search involves the solution of the 
time-dependent PDEs for the model for a specific value of $A_3$, with all other parameters held 
fixed, and the criterion by which we adjust the bisection bracket is whether or not the simulation 
results in black hole formation.)

A series of snapshots of $\pa M(t,r)/\pa r$ (where $M(t,r)$ is the mass aspect function) for a marginally 
subcritical evolution is shown in Fig.~\ref{dmdr}.
Full analysis of the results of this simulation indicate that the boson star enters what
we identify as the critical
state at $t \approx 130$, and remains in that state until $t \approx
510$.  It is worth noting that the boson star actually completes its collapse into a more 
compact configuration well after the real scalar field has dispersed from the boson star 
region.   We also note that the amount of time, $\tau$, spent in the critical 
state---$\tau \approx 380$ in this case---is a function of how closely the control parameter
has been tuned to criticality.  Specifically, we expect $\tau$ to be linear in $\ln|A_3 -
A_3^\star|$
(see (\ref{tau-scaling})), and we will display evidence for this type of scaling below.

\begin{figure}
  \begin{center}
\includegraphics[width=8.0cm,clip=true]{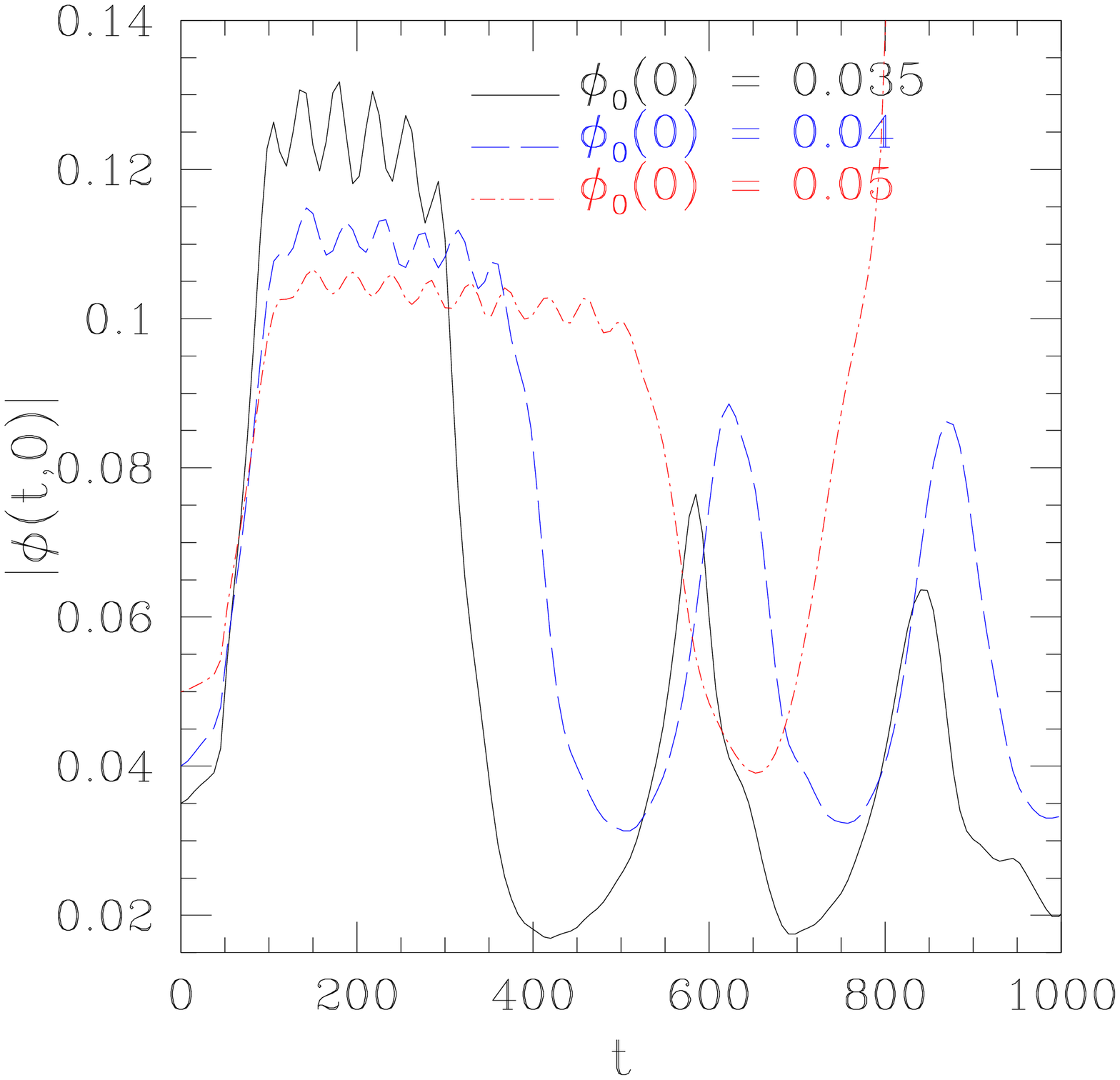}
    \caption
[Time evolution of central value of the modulus of scalar field for subcritical evolution of 
perturbed boson stars]
{Time evolution of the central value of the scalar field modulus for subcritical evolution of perturbed 
boson stars.
The figure shows the time evolution of $|\ph(t,0)|$ for marginally subcritical evolutions 
generated from boson star
initial states with $\ph_0(0) = 0.035, 0.04$ and
$0.05$.  See the text for a description of key features of this plot.
} 
    \label{modph}
  \end{center}
\end{figure}

%
%

%
%
\begin{figure}
  \begin{center}
\includegraphics[width=8.0cm,clip=true]{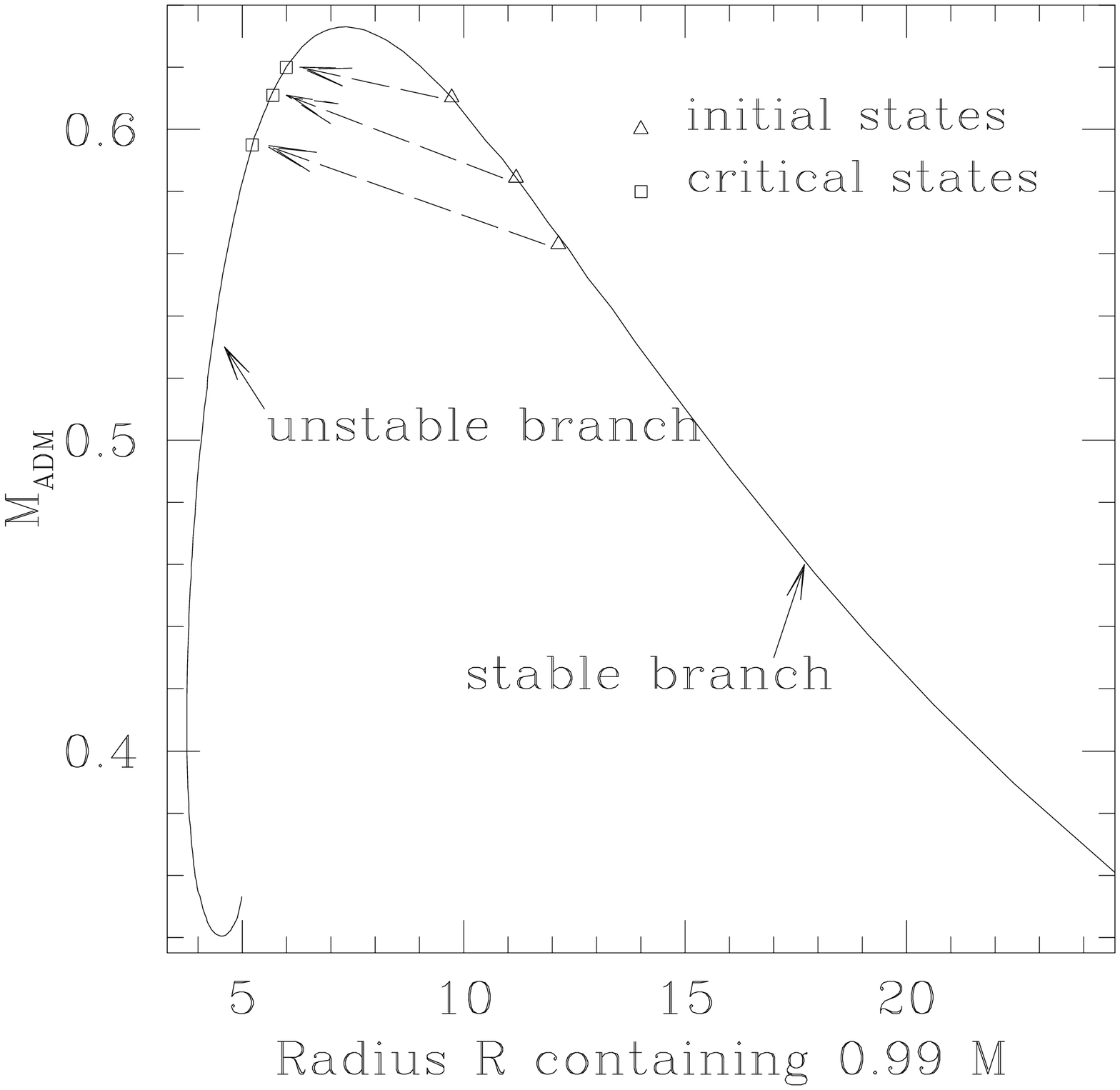}
    \caption
[Transition of perturbed boson stars in critical evolutions]
{Transition of perturbed boson stars in critical evolutions.  The solid curve shows the 
parametric mass {\em vs} radius plot of static boson stars 
(curve parameter $\phi_0(0)$ increasing from right to left), where we 
have defined the stellar radius, $R$, so that $M(R) = 0.99 \, M(\infty) = 0.99 \, M_{\rm ADM}$.
Triangles label the initial configurations, squares show the corresponding critical 
solutions (identified as one-mode-unstable boson stars with 
oscillations that are largely in the fundamental mode), 
and the dashed arrows represent schematically the transition between the initial and critical 
states. See the text for more details.}
    \label{transit}
  \end{center}
\end{figure}

Fig.~\ref{modph} shows the time evolution of the central modulus of the complex
scalar field for marginally subcritical evolutions generated from boson star 
initial states with $\ph_0(0) = 0.035, 0.04$ and
$0.05$.  From the figure we can see that in all three cases the perturbed stars enter an excited, critical 
state at $t\approx 100$ and remain in that state for a finite time which is a function of $\phi_0(0)$
(i.e.\ of the initial state).  Additionally, at least for the cases $\phi_0(0) = 0.035$, $\phi_0(0) = 0.04$,
the figure provides evidence that following the critical evolution phase, the excited stars relax
to states characterized by large amplitude oscillations of the complex field.  This behaviour will be 
examined in more detail below.  Finally, also apparent in the plot are the smaller-amplitude 
oscillations that occur
during the periods of  critical evolution. Previous work~\cite{shawley:phd,scott_matt:2000} indicated
that these oscillations can be interpreted as excitations of the (stable) first {\em harmonic} mode of 
the unstable boson star that is acting as the critical solution---the 
unstable {\em fundamental} mode is the one that determines whether or not the configuration will
evolve to a black hole.
Although we have not studied this matter in any detail, we assume that the same picture holds for 
our current calculations.

The results from our simulations of critically perturbed boson stars are thus in agreement with 
the previous studies~\cite{shawley:phd,scott_matt:2000} which identified the critical states as 
excited (primarily in the first harmonic mode), unstable boson stars.  Following that work
we display in Fig.~\ref{transit} an approximate correspondence between
the initial boson stars and the critical solutions.
The solid line traces the one-parameter family of static boson stars (parameterized as usual by
$\phi_0(0)$), where we have defined the radius, $R$, of a boson star so that 
$M(R) = 0.99 \, M(\infty)  = 0.99 \, M_{\rm ADM}$.
The triangles indicate the initial stable boson star configurations, the squares indicate our 
best estimate of the corresponding unstable critical boson star states, and each arrow schematically 
depicts the transition between the two states that is induced by the perturbing scalar field. 
We note that to identify which unstable boson star is acting as the critical solution---which
is equivalent to identifying an effective value of $\phi_0(0)$---we time average 
the central modulus of the complex field, $\vert \phi(t,0) \vert$ during the period of critical evolution.
In addition, in accord with previous results, we observe that in all cases the mass of the unstable 
critical state is {\em larger} than that of the progenitor boson star, indicating that a 
significant amount of mass-energy is extracted from the massless scalar field through its purely 
gravitational interaction with the complex field.

\begin{figure}
  \begin{center}
\includegraphics[width=8.0cm,clip=true]{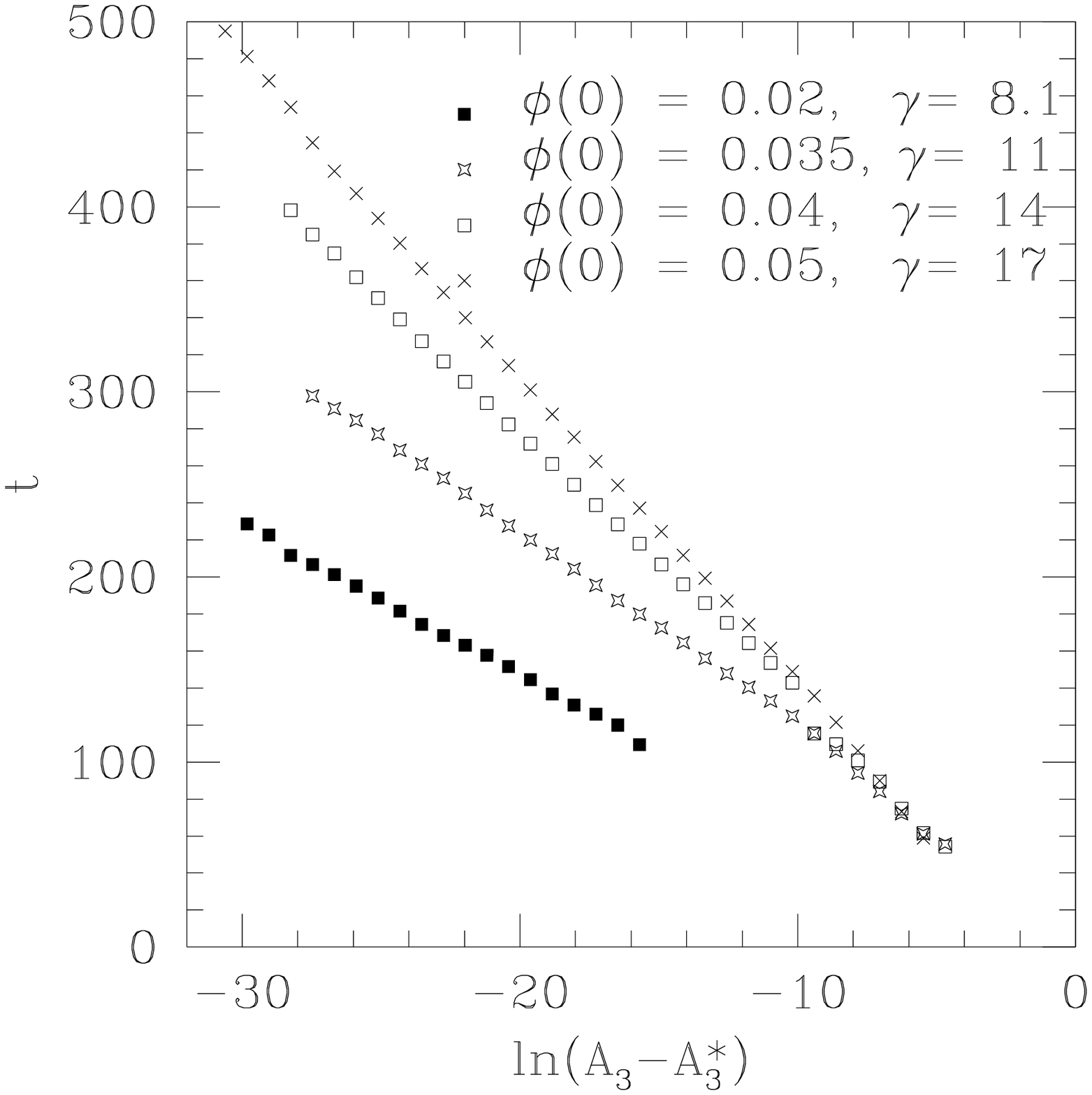}
    \caption[Measured lifetime scaling laws for critically perturbed boson stars]
{Measured lifetime scaling laws for critically perturbed boson stars.
This figure shows the measured lifetimes of various near-critical evolutions of 
perturbed boson stars as a function of $\ln|A_3 - A_3^\star|$, for cases with 
$\ph(0) = 0.02, 0.035, 0.04$ and $0.05$.  Quoted scaling exponents, $\ga$ 
(see~(\ref{scaling-2})), are computed from linear least-squares fits to the data.
The apparent convergence of the data for different $\ph_0(0)$ as $\ln\vert A_3-A_3^\star\vert \to 0$
is not significant, as it reflects calculations {\em  far} from criticality i.e.\ far from
the $\ln\vert A_3 - A_3^\star\vert \to -\infty$ limit. See the text for additional details.
}
    \label{t_vs_lndp}
  \end{center}
\end{figure}

As discussed previously, for both subcritical and supercritical simulations,
the closer one tunes $A_3$ to the critical value $A_3^\star$, the longer the 
perturbed star will persist in the critical state.  Specifically, we observe scaling 
of the lifetime, $\tau$, of the critical evolution of the form
\beq
\label{scaling-2}
   \tau(A_3) \sim -\gamma \ln | A_3 - A_3^\star | \,,
\eeq
where we define the lifetime to be the lapse of coordinate time from the start 
of the evolution, $t=0$, to the time of first detection of an apparent horizon, and where $\gamma$ is 
a scaling exponent that depends on which of the infinitely many one-mode unstable boson stars acts as 
the critical solution in the particular scenario being simulated.  We note that 
the details of the definition of $\tau$ are not important to the determination of 
$\gamma$ in~(\ref{scaling-2}) since $\gamma$ actually measures the {\em differential} in lifetime 
with respect to changes in $A_3 - A_3^\star$, and this differential is insensitive 
to precisely how we define $\tau$,
at least as $A_3 \to A_3^\star$.  In addition, we note that in using coordinate time in our definition
of the scaling relationship~(\ref{scaling-2}), we are defining the scaling with respect to proper time 
at spatial infinity.  Another choice---arguably more natural---would be to define $\tau$ in terms of 
the proper time measured by an observer at rest at $r=0$ (central proper time).  Since the critical 
solutions are nearly static, the relation between these two different definitions of time would be 
a specific factor for each distinct value of $\phi_0(0)$, and would thus 
lead to a $\phi_0(0)$-dependent 
``renormalization'' of the scaling exponents, $\gamma$.

Fig.~\ref{t_vs_lndp} shows measured scaling laws from supercritical evolutions of 
perturbed boson stars  defined by $\ph_0(0) = 0.02, 0.035, 0.04$ and 
$0.05$.  It is clear from
these plots that, at least as $A_3 \to A_3^\star$, we have lifetime scaling of the 
form~(\ref{scaling-2}).  Estimated values of $\gamma$---computed from linear least-squares 
fits to the plotted data---are $\ga = 8.1, 11, 14, 17$ for $\ph_0(0) = 0.02, 0.035, 0.04, 
0.05$, respectively.  We note that according to the now standard picture of critical 
collapse (see for example~\cite{gundlach:2003}), each value of $\ga$ can
be identified with the reciprocal Lyapunov exponent (i.e.\ growth factors) of the 
single unstable mode associated with the corresponding critical solution.  Again, the reason
that we observe different values of $\ga$ for different choices of initial boson star (different 
values of $\phi_0(0)$) is that distinct critical solutions are being generated in the 
various cases.  That is, we cannot expect universality (with respect to initial data) in this case 
because the model admits an entire family of one-mode unstable solutions that sit at the threshold 
of black hole formation.

\subsection{Final Fate of Subcritical Evolutions}
\lab{sec:latetime}
In previous work on the problem of critically perturbed spherically symmetric 
boson stars~\cite{shawley:phd,scott_matt:2000}, it was conjectured that the 
end state of subcritical evolution was characterized by {\em dispersal} of the boson 
star to large distances (relative to the size of the initial, stable star).
This conjecture was at least partially influenced by the behaviour observed, for 
example, in the collapse of a {\em massless} scalar field~\cite{choptuik}, where 
subcritical evolutions {\em do} involve complete dispersal of the field.  However, 
another key reason for what we claim is a misidentification of the true subcritical
end-state, was that the simulations described in~\cite{shawley:phd,scott_matt:2000}
simply were not carried out for sufficient coordinate time to 
observe the nature of the late-time dynamics.
Our current simulations strongly suggest that subcritical evolutions
lead to a ``relaxation'' of the critically perturbed state to something that 
approximates a boson star (not necessarily the original star) undergoing large amplitude 
oscillations. As argued in the next subsection, these oscillations
can largely be identified with the fundamental perturbative mode 
associated with the final boson star state. The numerical evidence also suggests that,
at least in many cases, these
oscillating configurations eventually {\em  re-collapse} and form black holes; 
a ``prompt" re-collapse can be seen in the $\ph_0(0) = 0.05$ data in Fig.~\ref{modph}.

\begin{figure}
\begin{center}
\includegraphics[width=8.75cm,clip=true]{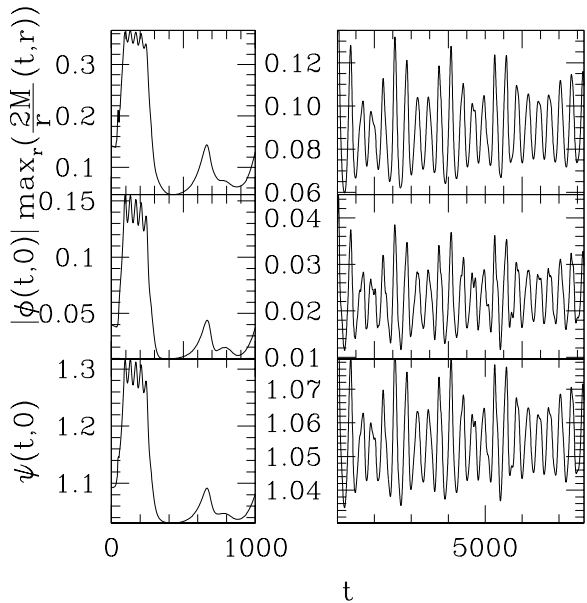}
\caption
[Long time behaviour of subcritical evolution for $\ph(0,0) = 0.04$]
    {Long time behaviour of subcritical evolution for $\ph(0,0) = 0.04$ with $r_{\rm max}=200$. 
This figure shows the long-time behaviour of $\max_r(2M(t,r)/r)$, $\vert \phi(t,0)\vert$ and $\psi(t,0)$ for 
a near-critically perturbed boson star ($\ph_0(0) = 0.04, A_3^\star \approx 0.00342$).
The left side of the figure shows the evolution of the perturbed star in its critical
state ($100 \lesssim t \lesssim 300$), and the evolution shortly after the star leaves its
critical state.  The right side of the figure focuses on oscillations seen at  later times
$1000 \le t \le 7680$.
These plots provides evidence that the final state of subcritical evolution is 
characterized by large amplitude oscillations about something approximating a boson
star on the stable branch, rather than dispersal of the complex field as suggested
in~\cite{shawley:phd,scott_matt:2000}.
Detailed calculation (see Sec.~\ref{perttheory}) shows that the 
pulsation frequency is approximately the fundamental mode frequency 
computed from perturbation theory about a background stable boson star solution
with $\ph_0(0)=0.023$.
We also note the overall lower-frequency modulation of the post-critical oscillations. 
This effect is not yet understood, although one possible explanation---namely that the 
envelope modulation represents ``beating'' of the fundamental and first harmonic modes---appears to 
be ruled out. 
}
    \label{tmrphps}
  \end{center}
\end{figure}

Fig.~\ref{tmrphps} displays the long-time behaviour of $\mbox{max}_r(2M(t,r)/r)$, $\vert
\phi(t,0)\vert$ and 
$\psi(t,0)$ for a near-critically perturbed boson star ($\ph_0(0) = 0.04, A_3^\star \approx 0.00342$)
for $r_{\rm max}= 200$ (with mesh spacing $\De r = 200/4096 \approx 0.049$).  
Note that this is a {\em subcritical} evolution, so that a black hole does {\em not} form.
As shown in more detail in previous figures, the boson star enters a critical
state (well approximated by an unstable boson star) shortly after the real scalar field leaves 
the computational domain ($t\approx100)$.
While in the critical state, the star oscillates with what we assume is the frequency of the 
first harmonic, as
computed from perturbation theory using the unstable boson star state as the 
background (see \cite{shawley:phd,scott_matt:2000}).  At $t\approx 300$ the star 
begins to evolve away from
the more compact critical configuration, decreases in central density, expands
in size, and starts to pulsate with a different frequency.  Although at late
time the oscillation amplitudes are much larger than those seen in the critical phase of 
evolution, we will show in the following section that the oscillations can nonetheless be largely 
attributed to excitations of the fundamental perturbative mode
associated with the final boson star state.

%
%

\begin{figure}
\begin{center}
\includegraphics[width=8.0cm,clip=true]{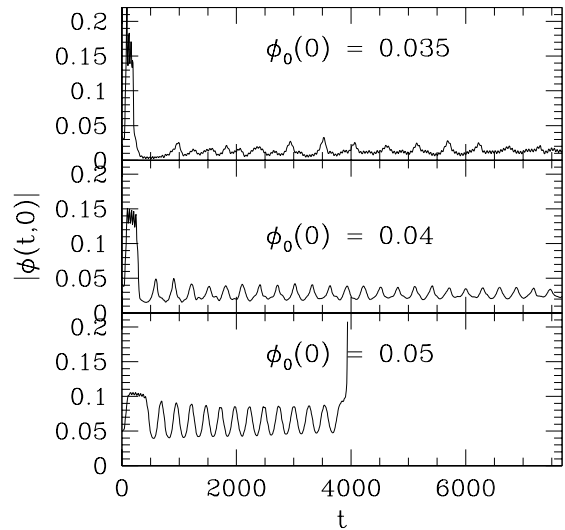}
\caption
[]
{Long time behaviour of subcritical evolution with initial configurations $\ph_0(0) =
0.035, 0.04$ and $0.05$, for $r_{\rm max} = 100$.  
The figures show the modulus of the central scalar field values, $\vert \phi(t,0)\vert$, 
{\em vs} time, using 
the same resolution $\De r = 50/1024$ used to generate the data shown in Fig.~\ref{tmrphps}.  
Each of the three distinct 
boson stars is driven to a different critical solution, and subsequently 
relaxes to a different final oscillatory state.  
This provides evidence that the final end state of
marginally subcritical evolution in generic driven boson stars does {\em not} involve dispersal of the 
bulk of the complex field to infinity.
}
    \label{modphrmax100}
  \end{center}
\end{figure}

Fig.~\ref{modphrmax100} shows the long-time behaviour of the modulus of the central
value of scalar field, $|\ph(t,0)|$, for initial configurations with 
$\ph_0(0) = 0.035, 0.04$ and
$0.05$, with $r_{\rm max} = 100$, but with 
$\De r$ maintained at $50/1024$ as in Fig.~\ref{tmrphps}.
Again, we use $A_3$ to tune the evolution of the boson stars to criticality and the figure 
shows a marginally subcritical evolution.
In general, the computed value of $A^\star_3$ is a function of $r_{\rm max}$, as is the specific
stable boson star to which the critical evolution relaxes. 
However, the results shown in the figure support our claim that an oscillatory phase (rather 
than dispersal) {\em generically} follows near-critical evolution of driven boson stars in the 
marginally subcritical case.

%
%
%
\begin{figure}
\begin{center}
\includegraphics[width=8.0cm,clip=true]{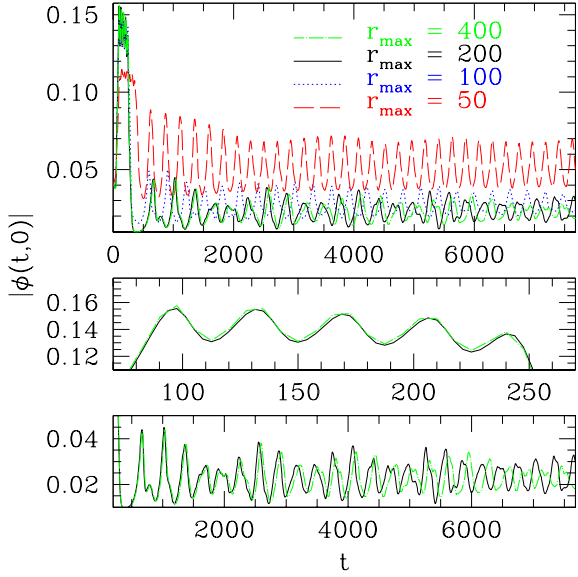}
\caption
{Long time behaviour of subcritical evolution with an initial boson star characterized by
$\ph_0(0) = 0.04$, for $r_{\rm max} =50, 100, 200$ and $400$.
The figures show the modulus of the central scalar field values, $\vert \phi(t,0)\vert$, 
{\em vs} time, with
the resolution $\De r$ fixed at $50/1024$ as in previous figures.  The evolutions
are tuned to criticality for different $r_{\rm max}$ (see Table~I).
The top figure shows the overall evolutions for $r_{\max}=50,100,200$ and $400$ from $t=0$ to $t=7680$.  The
middle figure focuses on the evolution of the perturbed boson star during the period 
of near-critical evolution, $70\le t\le 270$, for the cases $r_{\rm max} = 200$ and $400$.
The near coincidence of the two curves in this case provides strong evidence for convergence of our 
calculations (at fixed spatial resolution) as $r_{\rm max}\to\infty$ .  The bottom figure
focuses on the late time evolution---$200\le t\le 7680$---again for $r_{\max}=200$ and $400$,
and provides additional support for our claim that 
the final oscillatory states we observe in subcritical evolution are not an artifact
of the use of a finite computational domain.}
    \label{modph.04}
  \end{center}
\end{figure}

\begin{figure}
\begin{center}
\includegraphics[width=8.0cm,clip=true]{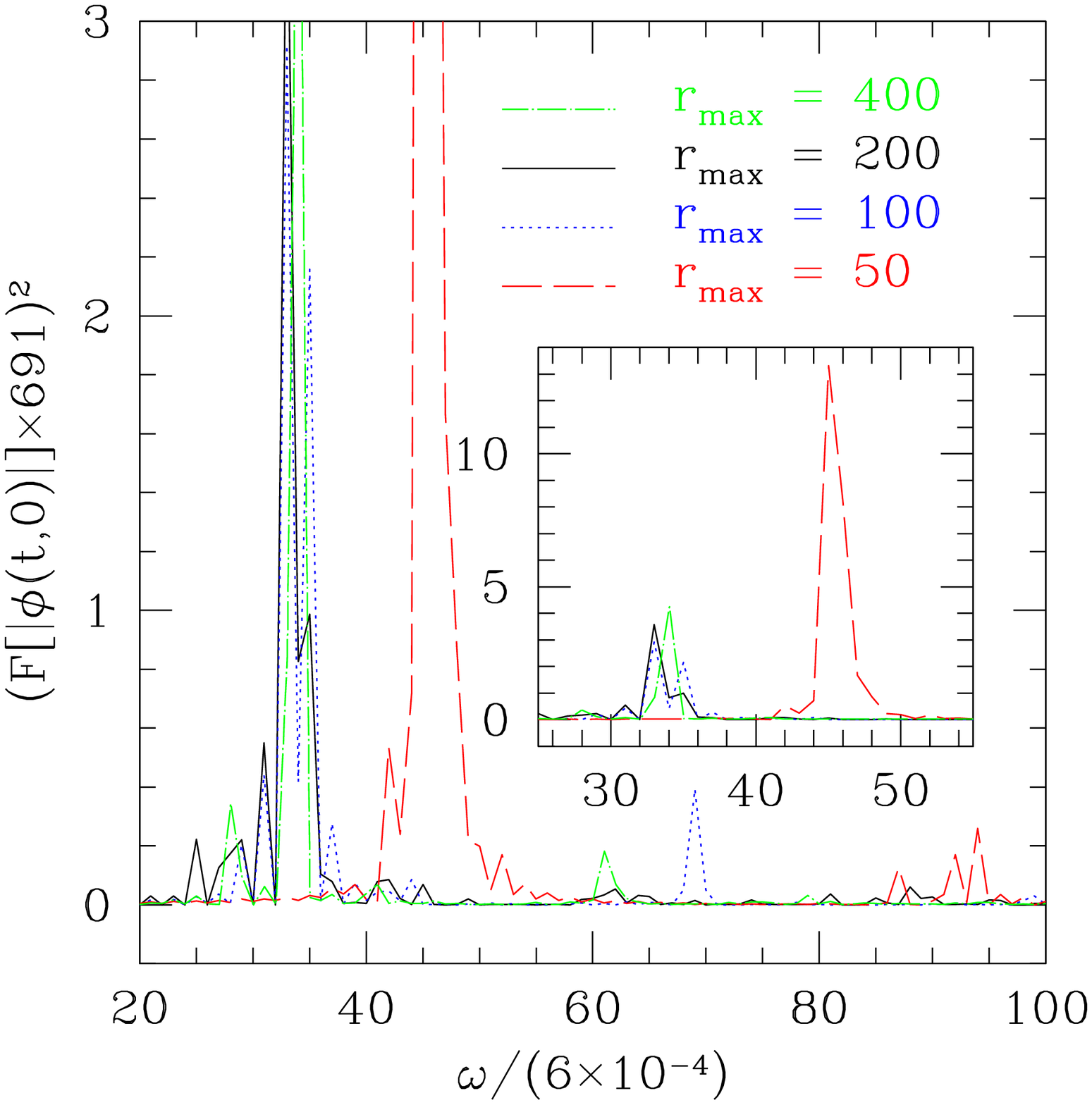}
\caption
{Long time behaviour of subcritical evolution with an initial boson star characterized by $\ph_0(0) =
0.04$, for $r_{\rm max} =50, 100, 200$ and $400$.  
The figures show the square of the (discrete) Fourier transform 
$\mathcal{F}\left[ \vert \phi(t,0)\vert\right]$, of the central scalar field modulus, 
using the same calculations described in Fig.~\ref{modph.04}.
The transform is taken from a data set defined at 691 discrete times, $t^n$  satisfying
$2500\lesssim t^n\lesssim 7700$, during which time 
the critically perturbed boson star is in its final oscillatory state.
Again, the resolution, $\De r=50/1024$, is the same used in previous calculations
The fundamental mode computed for the case $r_{\rm max}=200$ is approximately 
$\om \approx 33\times 6 \times 10^{-4} = 0.0198$, 
in good agreement with our perturbation-theory estimate computed in Sec.~\ref{perttheory}.
The figure shows that the computed frequency of the fundamental mode converges for 
increasing $r_{\rm max}$.
The graph also shows evidence for at least one higher overtone which persists as
$r_{\rm max}\to\infty$.  The figure inset shows the overall amplitudes of the computed Fourier
components.}
    \label{fft.04}
  \end{center}
\end{figure}

Fig.~\ref{modph.04}
shows the long time behaviour of subcritical evolution of the modulus of 
the central scalar field value, $\vert\ph(t,0)\vert$, with an initial boson star given by
$\ph_0(0) = 0.04$. Here, we vary the position of the outer edge of the computational domain
$r_{\rm max}$, while keeping the resolution, $\De r$, fixed at $50/1024$ 
as previously.  For each of $r_{\rm max} = 50, 100, 200$ and $400$, we tune $A_3$ to
generate a critical evolution (the specific values of $A^\star_3$ obtained are 
listed in Table~I).  This set of calculations provides evidence for the 
convergence of the critical solution (including the critical value of the control parameter, $A_3$),
as $r_{\rm max} \to \infty$ at fixed resolution.  This in turn strongly suggests that the 
final oscillatory states identified in subcritical evolutions are not artifacts of our 
use of a finite computational domain. 

In order to illuminate the nature of typical post-critical oscillations,
Fig.~\ref{fft.04} shows the square of the
discrete fast Fourier transform, $\mathcal{F}\left[ \vert \phi(t,0)\vert\right]$, of
the central scalar field modulus for the same set of simulations used to prepare 
Fig.~\ref{modph.04}.  The transform is taken for discrete times, $t^n$, satisfying
$2500 \lesssim t^n \lesssim 7000$, a period when the boson star has undergone the 
transition from critical evolution to post-critical oscillation.
The figure clearly shows the convergence of the
fundamental mode oscillation, as well as a first harmonic.  The next section
provides a more detailed analysis of the observed fundamental mode excitations.

\subsection{Perturbation Analysis of Subcritical Oscillations} \lab{perttheory}
We now proceed to an application of perturbation theory to the oscillations seen in 
long-time evolutions of marginally subcritical configurations, such as those shown in 
Fig.~\ref{tmrphps}.
Here we follow~\cite{glesier_watkins:1989} and \cite{scott_matt:2000},
and refer the interested readers to those sources for details of the approach 
that we do no include here.  In particular, we emphasize that we have {\em not} carried out 
the complete perturbation analysis ourselves, but are simply using a computer code provided by
Hawley~\cite{hawley:private} to
analyze our current simulations.  Nonetheless, to make contact between the perturbative and 
simulation results, it is useful to briefly review the setup of the perturbative problem.

To formulate the equations for the perturbation analysis, we first rewrite the 
complex scalar field as
\beq
 \ph(t,r) = \left( \ps_1(t,r)+i \ps_2(t,r)\right) e^{-i \om t}\,,
\eeq
\noi
(Note that this representation is distinct from $\ph = \ph_1 + i \ph_2$, and the 
reader should be careful not to confuse the $\ps$'s used here with the conformal metric variable, $\ps$.)
Additionally, the spacetime metric is written in Schwarzschild-like (polar-areal) 
coordinates:
\bea
 ds^2 = -e^{\nu(t,r)}dt^2 &+& e^{\la(t,r)} dr^2 \nonumber \\ 
  &+& r^2\left(d \te^2+\sin^2 \te d
\vp^2\right)\,.
\eea

We further introduce four perturbation fields, $\de \la(t,r), \de \nu(t,r), \de \ps_1(t,r)$ and
$\de \ps_2(t,r)$, which represent the perturbations about the
equilibrium values $\la_0(r), \nu_0(r), \ph_0(r)$:
\bea
\la(t,r) &=& \la_0(r)+\de \la(t,r)\,,\\
\nu(t,r) &=& \nu_0(r)+\de \nu(t,r)\,,\\
\ps_1(t,r) &=& \ph_0(r) \left( 1+\de \ps_1(t,r)\right)\,,\\
\ps_2(t,r) &=& \ph_0(r) \de \ps_2(t,r) \,.
\eea
\noi

With the above definitions we can write the coupled  Einstein-Klein-Gordon field
equations as a set of PDEs for the functions $\de \la, \de \nu, \de \psi_1$ and $\de \psi_2$.
With some manipulation we can then eliminate $\de \nu$ and $\de \ps_2$ to produce 
a system of two coupled second-order PDEs for $\de \ps_1$ and $\de \la$:
\bea \nn
  \de {\ps_1}'' &=& - \left( \fr{2}{r} + \fr{{\nu_0}'-{\la_0}'}{2}\right) \de
{\ps_1}' -
\fr{\de \la'}{r {\ph_0}^2} + e^{\la_0-\nu_0} \ddot{\de \ps_1} \\ \nn
&& - \left[
\fr{{\ph_0}'}{\ph_0} \left( \fr{{\nu_0}'-{\la_0}'}{2} + \fr{1}{r}\right) +
\left( \fr{{\ph_0}'}{\ph_0}\right)^2  \right. \nn \\ 
 && \quad\quad \left. + \fr{1-r {\la_0}'}{r^2 {\ph_0}^2 } +
e^{\la_0-\nu_0} \om^2 - e^{\la_0} \right] \de \la \nn \\ 
&& + \, 2 e^{\la_0} \Bigg[ 1 + e^{-\nu_0} \om^2 \Bigg. \nn \\ 
&& \quad\quad\quad\quad \left. + \, e^{-\la_0} \left(
\fr{{\ph_0}'}{\ph_0}\right)^2 + r \ph_0 {\ph_0}'\right] \de \ps_1\,,
\eea

\bea \nn
  \de \la'' &=& -\fr{3}{2} \left( {\nu_0}' - {\la_0}'\right) \de \la' +
\Bigg[ 4 {{\ph_0}'}^2 + {\la_0}'' \Bigg. \nn \\
  &&\quad \left. + \, \fr{2}{r^2} - \fr{ ({\nu_0}' - {\la_0}')^2}{
2} - \fr{ 2 {\nu_0}' + {\la_0}'}{r} \right] \de \la\nn \\ 
&& + e^{\nu_0-\nu_0} \ddot{\de \la} - 4 \left( 2 \ph_0 {\ph_0}' - r e^{\la_0}
{\ph_0}^2\right) \de {\ps_1}' \nn \\
&& - 4 \Bigg[ 2 {{\ph_0}'}^2 - r e^{\la_0} {\ph_0}^2 \Bigg. \nn \\ 
 && \quad \quad \times \left. \left( 2
\fr{{\ph_0}'}{\ph_0} + \fr{2 {\nu_0}' + {\la_0}'}{2}\right) \right] \de \ps_1\,.
\eea

\noi
Note that these equations involve only second time derivatives (i.e.\ there are 
no terms involving $\dot{\delta \psi_1}$ or $\dot{\delta \lambda}$), and that they 
are {\em linear} in the second time derivatives.   If we thus assume 
a harmonic time-dependence for the perturbed fields:
\bea \lab{harmonicdep1}
 \de \ps_1(t,r) &=& \de \ps_1(r) e^{i \si t}\,,\\ \lab{harmonicdep2}
 \de \la_1(t,r) &=& \de \la_1(r) e^{i \si t}\,,
\eea
\noi
then the equations for the perturbations contain $\si$ only in the form 
$\si^2$, and the sign of 
$\si^2$, as computed by solving a particular mode equation, determines the stability of that
mode.  (Note that the system can be shown to be self-adjoint so that the
values of $\si^2$ must be real.)  If any of the values of $\si^2$ 
are found to be negative, then the
associated perturbations will grow and the boson star will be unstable.  Moreover, as
the eigenvalues form an infinite discrete ordered sequence, examining the 
fundamental radial mode ${\si_0}^2$ determines the overall stability of any particular 
star with respect to radial perturbations.

In order to compare the simulation results with those given by perturbation
theory, we first observe
that there is a difference in the choice of the time coordinates used in the two calculations.
Specifically, 
in the perturbative analysis~\cite{glesier_watkins:1989,scott_matt:2000}, the lapse 
is chosen to be unity at the origin, so we have
\[
\sigma^2 \Bigl.\Bigr\vert_{\rm perturbative} \rightarrow
\frac{\sigma^2}{\alpha^2} \Bigl.\Bigr\vert_{\rm simulation}\,.
\]

\noi
We also note that there is a factor of 2 difference in the definitions of $T_{\mu
\nu}$ used in the two calculations, and that the 
definition of the complex field, $\ph(t,r)$, in the perturbative calculation includes a factor of $\sr{8 \pi}$.  We thus 
have
\[
\ph\Bigl.\Bigr\vert_{\rm perturbative} \rightarrow
\sr{4 \pi}\ph\Bigl.\Bigr\vert_{\rm simulation}\,.
\]

\noi
The numerical technique for obtaining the fundamental mode and first harmonic
mode frequencies of boson stars has already been described in \cite{scott_matt:2000}
and will not be repeated here; again, we will simply quote and use results from that 
study.
From Fig.~\ref{tmrphps} we note that
there are 10 oscillations between $t = 2553.8$ and $t=5583.8$, giving a period $T
\approx 333$.  Hence we have an oscillation frequency $\si = 2 \pi/T \approx 0.019$.  
The time average of the lapse function, $\langle \al(t,0) \rangle_t$, in the interval 
is 0.89, and so $\si^2/\al^2 \approx 0.00045$.  We also compute the time average of
$\ph(t,0)$ in the interval, and use the resulting value to identify the 
stable boson star solution about which we perform the 
perturbation analysis.  We find $\langle \ph_0(t,0) \rangle_t \approx
0.023 \times \sr{4 \pi} = 0.0815$.  For a boson star with $\ph_0(0) = 0.0815$, the perturbative 
calculations (see Fig.~7 of \cite{scott_matt:2000}) 
predict $\si_0^2 = 0.00047$, which is in reasonable agreement with the 
simulation results.  Hence the oscillations that occur in the post-critical
regime appear to be largely fundamental mode oscillations of a final-state, stable, boson star.  
We also remark that since the oscillations are of such large amplitude, it does not appear
possible to precisely identify an effective background state (i.e.\ an effective value 
of $\ph_0(0)$), so the level of agreement in the oscillation frequencies is probably 
as good as one could expect.

\section{Summary} 
\lab{sec:conclusions}

We have investigated type I critical phenomena of ground state boson stars
in maximal-isotropic coordinates by perturbing the stars with in-going pulses 
of a real scalar field.  
In particular, contrary to previous claims, we find that 
the end state of generic subcritical evolution is a stable boson star executing 
large amplitude oscillations, and that the oscillations can be largely 
understood as excitations of 
the fundamental normal mode of the end-state star.
For the particular example that we examined in detail, the  oscillation frequency
of the post-critical state was estimated to be 
$\si^2/\al^2 \approx 0.00045$, in good agreement
with the frequency of the fundamental mode computed in perturbation theory, 
$\si^2_0 = 0.00047$.

\subsection*{Acknowledgments} 
We would like to thank S. H. Hawley, who provided the perturbation code for
generating the fundamental and first harmonic modes for boson stars via
perturbation theory.
A part of the numerical computation was carried out on the
vn.physics.ubc.ca Beofwulf cluster which was funded by CIAR, CFI, NSERC and
NSF.

\appendix
\section{Finite difference Algorithm} \label{fda}
\lab{sec:FDA}

Here we present the details of the numerical method used in our computations.
We solve the PDEs 
(\ref{hamiltonian_constraint})--(\ref{isotropic_condition})
 by a finite difference method.  
We replace the
$(t,r)$ continuum by a discrete lattice of grid points, and
approximate the continuum field quantities
${\cal F}=\{\al, \bt, \ps, \krr, \ph_i, \Ph_i, \Pi_i\}$, where  $i=1,2,3$,
 by a set of grid 
functions ${\cal F}^h=\{\al^h, \bt^h, \ps^h, \krr^h, \ph_i^h, \Ph_i^h, \Pi_i^h\}$ which are solutions of the finite
difference approximation (FDA) of
the PDEs.  
Denoting the uniform (constant) spatial and temporal mesh spacings by $\De r$ and $\De t$,
respectively, the finite difference grid is given by $(t^n,r_j)$ 
where 
$ r_j = r_0 + (j - 1) \De r $, $j = 1, \cdots, N_r$ and $t^n = n \De t$,
$n = 0, \cdots, N_t$.  For any grid function $u^h \in {\cal F^h}$, the value at $(t^n,r_j)$
is denoted by $u^n_j $ and is an approximation
of the continuum value $u(t^n,r_j)$.

\begin{figure}
  \begin{center}
\includegraphics[width=8.0cm,clip=true]{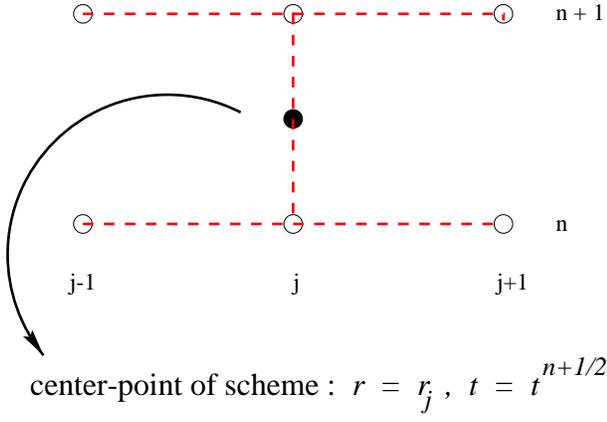}
    \caption[Stencil for $O(h^2)$ Crank-Nicholson approximation]
{
Stencil for an $O(h^2)$ Crank-Nicholson scheme for a PDE in one space dimension and time.
}
    \label{cn}
  \end{center}
\end{figure}

In discretizing evolution equations 
(\ref{klein_gordon_eq1})$-$(\ref{klein_gordon_eq3}) we make exclusive use of 
Crank-Nicholson schemes, with second order spatial differences.
The key idea of a Crank-Nicholson method is to keep the differencing {\em centred} in time 
as well as in space, and a typical
stencil used for such a scheme is illustrated in Fig.~\ref{cn}. 
The constraint equations (\ref{hamiltonian_constraint}) and
(\ref{momentum_constraint}) are coupled, nonlinear, ordinary differential equations, and 
following $O(\De r^2)$ finite differencing (see below) 
and are solved using a point-wise Newton's method.  That is, at each grid point,
$(t^n,r_j)$, we solve for the pair $(\psi^n_j,{(K^r{}_r{})}^n_j)$ using Newton's method for two equations
in two unknowns. 
The slicing condition 
(\ref{maximal_condition}) is linear, so, after being discretized using second-order 
finite differences, can be solved directly using a 
tridiagonal solver.  Finally, once the values $\al^n_j$ and ${K^r{}_r{}}^n_j$ have been computed,
an $O(\De r^2)$ discretization of~(\ref{isotropic_condition}) is easily integrated to yield
the $\beta^n_j$. 

To aid in the presentation of the finite
difference equations, it is convenient to define the following
difference operators:

%
%

\[
\Dtp u\nj = 
\fr{ u\npoj - u\nj }{\De t} \,,
\]

\[ 
\Drz u\nj = 
\fr{u\njpo - u\njmo}{2 \De r} \,,
\]

\[
\Drzb  u\nj = \fr{3 u\nj - 4 u\njmo + u\njmt}{2 \De r} \,,
\]

\[
\De^{^{\scr r}}_{_{\scr \pm}} u\nj = 
\fr{ \pm u^{^{\scr n}}_{_{\scr j \pm 1}} \mp
 u\nj }{ \De r} \,,
\]

\[ 
\Dhrz u\nj = \fr{ u\njph - u\njmh }{\De r} \,,
\]

\[
\Drsz u\nj = \fr{  u\njpo - u\njmo }{
r_{_{\scr j+1}}\,^2 - r_{_{\scr j-1}}\,^2 } \,,
\]

\[
\Drcz u\nj = \fr{  u\njpo - u\njmo }{
r_{_{\scr j+1}}\,^3 - r_{_{\scr j-1}}\,^3} \,.
\]

\[
\Dhrcz u\nj = 
\fr{ u\njph - u\njmh}{
r_{_{\scr j+\ha}}\,^3 - r_{_{\scr j-\ha}}\,^3} \,,
\]
\noi
and the averaging operator

\[
\mu^{^{\scr t}}_{_{\scr \pm}} u\nj =  \ha \left( u^{^{\scr
 n \pm 1}}_{_{\scr j}} + u\nj
 \right),
\]

\[
\mu^{^{\scr r}}_{_{\scr \pm}} u\nj = \ha \left( u^{^{\scr
 n}}_{_{\scr j \pm 1}} + u\nj
 \right)\,.
\]

\noi
We also define $\bar{\mu}^{^{\scr r}}_{_{\scr \pm}}$, which has the same
definition as $\mu^{^{\scr r}}_{_{\scr \pm}}$, but which has a higher precedence
over other algebraic operations, e.g.,

\[
\bmurp \left( \fr{ f g^2}{h}\right)^n_j = \fr{(\bmurp f\nj) (\bmurp
{g\nj})^2}{\bmurp h\nj}\,.
\]

\noi
The FDAs of the Klein-Gordon equations can then be written as:

\beq
\Dtp (\ph_i)\nj = \mutp
\left( \fr{\al}{\ps^2} \Pi_i + \bt \Ph_i \right)^n_j \,,
\eeq

\beq \lab{fd_klein_gordon_eq_2}
 \Dtp (\Ph_i)\nj = \mutp \Drz
\left( \bt \Ph_i + \fr{\al}{\ps^2} \Pi_i\right)^n_j \,, 
\eeq

\bea 
\Dtp (\Pi_i)\nj & = & \mutp \Biggl\{ \fr{3}{(\ps^4)\nj}  \Drcz \left[ r^2 \ps^4 \left( \bt 
\Pi_i + \fr{ \al }{ \ps^2} \Ph_i\right)\right]^n_j \Biggr. \nn \\ 
 &-& \left.  \left[ \al \ps^2 m^2 \ph_i
\left( 1 - \de_{i3} \right)\right]^n_j \right. \nn \\
&-& \Biggl.    \Biggl[ (\al \krr)\nj + 2 \bt\nj \, \fr{\Drz (r
\ps^2)\nj }{(r \ps^2)\nj } \Biggr] (\Pi_i)\nj \Biggr\}
\,.
\lab{fd_klein_gordon_eq_3}
\eea

\noi
where $i =1,2,3$.

The FDA of the Hamiltonian constraint is

\bea
\fr{ 3}{(\ps\nj)^5} &&\!\!\!\!\!\!\!\!\! \Dhrcz 
\left( r_{_{\scr j}}^2 \, \Dhrz \ps\nj \right) + \fr{ 3}{16}
(\krr)\nj \,^2 \nn \\
&=& - \pi\left( \fr{ \sum_{i=1}^3 \left( \Ph^2_i + \Pi^2_i\right)  }{\ps^4} +
m^2 \sum_{i=1}^2 \ph_i^2 \right)^n_j \,,
\eea

\noi
and the FDA of the momentum constraint is 
\bea
\bmurm(\ps\nj)^2 &&\!\!\!\!\!\!\! \Drm (\krr)\nj 
 + 3\, \Drm (r \ps^2)\nj 
\, \bmurm \left( \fr{\krr}{r}\right)^n_j \nn \\
 &=& \bmurm \left[ -8
\pi \sum_{i=1}^3 \Pi_i  \Ph_i  \right]^n_j\,.
\eea

\noi
Similarly, the FDAs for the maximal-isotropic conditions are
\bea
\Drp \Drm \al\nj &+&
\fr{2}{(r \ps^2)\nj } \Drsz \left( r^2 \ps^2
\right)^n_j \Drz \al\nj \nn \\
 &+& \Biggl[ 4 \pi
m^2 \ps^4 \sum_{i=1}^2 \ph_i^2 \Biggr. \nn \\
&&\quad \Biggl. - \, 8 \pi \sum_{i=1}^3 \Pi_i^2 - \fr{3}{2}
(\ps^2 \krr)^2 \Biggr]^n_j \al^n_j  \nn \\
&=&  0 \,,
\eea

\noi
and

\beq
 r_{_{\scr j-\ha}}\, \Drm \left(
\fr{\bt}{r}\right)^n_j = \murm \left[\fr{3}{2} \al \krr \right]^n_j \, ,
\eeq

\noi
respectively, where  $r_{_{\scr j-\ha}} \eq (r_j + r_{j-1})/2$.

The regularity conditions are
implemented as
\bea
  \ps\no &=& \fr{4 \ps^{^{\scr n}}_{_{\scr 2}} - \ps^{^{\scr n}}_{_{\scr
3}}}{3} \,, \\
 \left({\krr}\right)\no &=& 0\,,\\
  \al\no &=& \fr{4 \al^{^{\scr n}}_{_{\scr 2}} - \al^{^{\scr n}}_{_{\scr
3}}}{3} \,,
\eea
\bea
 \mutp \left( (\ph_i)\no - \fr{4 (\ph_i)^{^{\scr n}}_{_{\scr 2}} - (\ph_i)^{^{\scr n}}_{_{\scr
3}}}{3}\right) &=& 0 \,, \\
 (\Ph_i)\no &=& 0\,,\\
 \mutp \left((\Pi_i)\no - \fr{4 (\Pi_i)^{^{\scr n}}_{_{\scr 2}} - (\Pi_i)^{^{\scr n}}_{_{\scr
3}}}{3}\right) &=& 0\,,
\eea
\noi
for all $i$ and $n$.  The outer boundary conditions are
\bea
\Dtp {\Ph_i}\nj  + \mutp \left( \Drzb {\Ph_i}\nj +
\fr{{\Ph_i}\nj}{r_j}\right) &=& 0\,, \\
\Dtp {\Pi_i}\nj  + \mutp \left( \Drzb {\Pi_i}\nj +
\fr{{\Pi_i}\nj}{r_j}\right) &=& 0\,.
\eea
\noi
We also adopt a scheme for 
numerical dissipation given by Kreiss and Oliger \cite{KO}.
In other words an additional term
\[
 \mutp \left( \Ddiss {\Ph_i}\nj \right)
\]

\noi
is added to the right hand side of (\ref{fd_klein_gordon_eq_2}), for $3 \le
j \le N_r-2$ (and
similarly to the right hand side of (\ref{fd_klein_gordon_eq_3}) for the $\Pi_i$), 
where $\Ddiss$ is defined by

\bea \lab{KOdiss}
 \Ddiss u\nj = -\fr{\ep_d}{16 \De t} && \!\!\!\!\!\! \left(  u\njpt - 4 u\njpo \right. \nn \\
   && \!\!\!\!\!\!\!\!\! \left. + \, 6 u\nj - 4 u\njmo + u\njmt \right) .
\eea
Here, $\epsilon_d$ is an adjustable parameter satisfying $0\le\epsilon_d<1$, and is typically
chosen to be 0.5.  We note that the addition of Kreiss-Oliger dissipation changes 
the truncation error of the FDAs at $O(\Delta t^3,\Delta r^3)$  and thus does 
{\em not} effect the leading order error of a second order ($O(\Delta t^2,\Delta r^2)$) scheme.
The dissipation is useful for damping high frequency solution components that are 
often associated with numerical instability.

\section{Convergence testing}
\lab{sec:convergencetest}

\begin{figure}
\begin{center}
\includegraphics[width=8.0cm,clip=true]{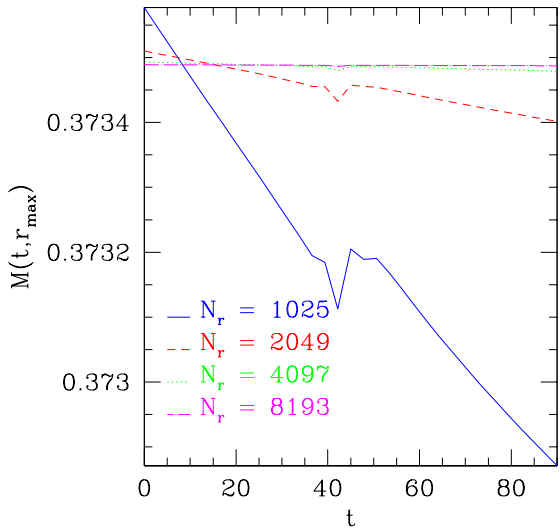}
    \caption[Convergence test of the spherically symmetric code]
{Convergence test of the spherically symmetric code.  The estimated
ADM mass, $M(t,r_{\rm max})$, is plotted against time, $t$, for four calculations 
using numbers of spatial grid points, $N_r$, of $1025, 2049, 4097$ and $8193$, so that 
the corresponding mesh spacings, $\Delta r$, are in a 8:4:2:1 ratio.
The initial data parameters for the
computations are: $\ph_0 = 0.01$ for the complex field, and $A_3=0.001, r_0 = 40$ and $\si
= 3$ for the massless field (see~(\ref{ph3-gaussian})).  
The mass decreases with time in general, with a significant fluctuation 
at $40 \leq t \leq 50$, when the real scalar field is close to the origin
and strongly interacts with the boson star.  The variation in the computed total mass tends 
to vanish as we go to higher resolution.  Combining results from the 
four calculations we find strong evidence that the finite difference scheme is second
order accurate as expected.}
    \label{admm_0123outer}
\end{center}
\end{figure}

Here we present the results of a convergence test of the code that evolves boson stars in 
spherical symmetry. 

In Fig.~\ref{admm_0123outer} we plot the mass aspect function
at the outer boundary of the computational domain, $M(t,r_{\rm max})$,
as a function of time, and from four computations with grid spacings,
$\Delta r$, in a $8$:$4$:$2$:$1$ ratio.   
As was the case for the calculations 
discussed in the main text Sec.~\ref{sec:results}, our convergence study 
uses a pulse of massless scalar field imploding onto a stable boson star 
So long as no scalar field (either real or complex) propagates off the 
computational grid, $M(t,r_{\rm max})$ should be constant in time (and equal to the ADM
mass), in the limit that $\De r \to 0$ (with $\De t \to 0$ implied since $\lambda$ is always held 
fixed as $\De r$ is varied).

In our test, the boson star has a central field value, $\ph_0 = 0.01$, while the incoming 
massless scalar field pulse is a gaussian of the form~(\ref{ph3-gaussian})
with $A_3 = 0.001, r_0 = 40$ and $\si = 3$.
The outer boundary is $r_{\rm max} = 300$,
and we compute with $N_r = 1025, 2049, 4097$ and $8193$.  
During the time interval
$40 \leq t \leq 50$, the real scalar field
is concentrated near the origin and interacts most strongly with the complex field. 
This results in a localized fluctuation of the computed ADM mass that is 
evident in the plots.  However, $M(t,r_{\rm max})$ clearly tends to a constant value 
as the resolution is increased. In addition, from the differences of $M(t,r_{\rm max})$
computed at different resolutions (e.g. $M^{\Delta r}(t,r_{\rm max}) - M^{2\Delta r}(t,r_{\rm max})$,
$M^{2\Delta r}(t,r_{\rm max}) - M^{4\Delta r}(t,r_{\rm max})$, etc.), we find strong
evidence that the overall difference scheme is converging in a second order fashion.

\bibliography{cbbs}
\end{document}